\documentclass[%
reprint,
twocolumns,
aps,
prl,
showpacs,
notitlepage,
nobibnotes,
nofootinbib,
superscriptaddress
]{revtex4-2}
\usepackage{booktabs}
\usepackage{float}
\usepackage{bigints}
\usepackage{caption}
\usepackage{psfrag}
\usepackage{grffile}
\usepackage{verbatim}
\usepackage{microtype}
\usepackage{multirow}
\usepackage{enumitem}
\usepackage{amsmath}
\usepackage{amssymb}
\usepackage{amsthm}
\usepackage{mathrsfs}
\usepackage{mathtools}
\usepackage{graphicx}
\usepackage{bbm}
\usepackage{subfig}
\usepackage[linesnumbered,ruled]{algorithm2e}
\usepackage{color}
\usepackage{lineno}
\usepackage{xcolor}

\definecolor{myblue}{rgb}{0.153,0.322,0.706}
\usepackage[colorlinks,linkcolor=myblue,urlcolor=myblue,citecolor=myblue]{hyperref}
\usepackage{geometry}
\newcommand{\be}{\begin{equation}}
\newcommand{\ee}{\end{equation}}
\newcommand{\ra}{\rightarrow}

\def\bc{\begin{center}}
\def\ec{\end{center}}
\def\bea{\begin{eqnarray}}
\def\eea{\end{eqnarray}}

\graphicspath{{./Figures}}

\definecolor{airforceblue}{rgb}{0.36, 0.54, 0.66}
\definecolor{brickred}{rgb}{0.8, 0.25, 0.33}
\definecolor{amber}{rgb}{1.0, 0.75, 0.0}
\definecolor{applegreen}{rgb}{0.55, 0.71, 0.0}
\definecolor{magenta}{rgb}{0.965, 0, 0.859}

\newcommand{\deletetext}[1]{\iffalse{{\color{red}{#1}}}\fi}
\newcommand{\newtext}[1]{{{#1}}}
\newcommand{\newtexttwo}[1]{{{#1}}}

\begin{document}

\title{Maximal dispersion of adaptive random walks} %based on large deviation theory}

\author{Gabriele Di Bona}
%\email{g.dibona@qmul.ac.uk}
\affiliation{School of Mathematical Sciences, Queen Mary University of London, London E1 4NS, England}

\author{Leonardo Di Gaetano}
%\email{di-gaetano\_leonardo@phd.ceu.edu}
\affiliation{Department of Network and Data Science, Central European University, 1100 Vienna, Austria}

\author{Vito Latora}
%\email{v.latora@qmul.ac.uk}
\affiliation{School of Mathematical Sciences, Queen Mary University of London, London E1 4NS, England}
\affiliation{Dipartimento di Fisica ed Astronomia, Universit\`a di Catania and INFN, I-95123 Catania, Italy}
%\affiliation{The Alan Turing Institute, The British Library, London NW1 2DB, England}
\affiliation{Complexity Science Hub Vienna (CSHV), Vienna, Austria}

\author{Francesco Coghi}
\email{francesco.coghi@su.se}
\affiliation{Nordita, KTH Royal Institute of Technology and Stockholm University, Hannes Alfvéns väg 12, SE-106 91 Stockholm, Sweden}

\date{\today}

\begin{abstract}
\deletetext{Random walks are the most versatile tool to explore a complex network. These include m}\newtext{M}aximum entropy random walks (MERWs)\deletetext{, which} are maximally dispersing and\deletetext{therefore} play a key role \newtext{in optimizing} \deletetext{as they optimize }information spreading \newtext{in various contexts}. However, building \deletetext{a}MERWs comes at the cost of knowing beforehand the global structure of the network\deletetext{ to be explored}\newtext{, a requirement that makes them totally inadequate in real-case scenarios}. Here, we propose an adaptive random walk (ARW), which instead maximizes dispersion by updating its transition rule on the local information collected while exploring the network. We show how to derive \deletetext{the}ARW via a large-deviation representation of \deletetext{the}MERW \deletetext{as a finite-time rare event of an unbiased random walk}and study its dynamics on synthetic and real-world networks.
\end{abstract}

\maketitle

%\tableofcontents

%%Intro

In the last decade we have assisted to a wave of new works studying extreme (rare) events associated to dynamical processes evolving on complex networks~\cite{Albeverio2006,Kishore2011,Kishore2012,Chen2014,Chen2015,Hindes2016,DeBacco2016,Hindes2017,Bianconi2018,Coghi2018,Staffeldt2019,Coghi2019,Gutierrez2021,Carugno2021,Kumar2020,Gupta2021,Gandhi}. The necessity to understand unlikely events comes from the fact that, although rare, their appearance determines the future of the system under study, which may be potentially catastrophic, e.g., earthquakes~\cite{Albeverio2006}. In this context, researchers have focused on random walks and their load and flow properties~\cite{Kishore2011,Kishore2012,Kumar2020,Gandhi} \deletetext{as }\newtext{for} models of traffic in transportation~\cite{Staffeldt2019,Gupta2021} and communication networks~\cite{Chen2014}, or on epidemic models and extinction events~\cite{Hindes2016}, or again on general order-disorder~\cite{Hindes2017} and percolation transitions~\cite{Bianconi2018,Coghi2018} to corroborate the robustness of networks. In these settings, rare events are often driven by internal noise\deletetext{ rather than environmental changes}, and their understanding could provide us with control mechanisms to keep away from harmful scenarios~\cite{Chen2015,Hindes2016,Hindes2017,Coghi2018}.

Among all dynamical processes evolving on networks, discrete-time biased and unbiased random walks (URWs) \deletetext{certainly }stand out as simple and insightful models of diffusion processes on discrete topologies~\cite{Noh2004}. 
\deletetext{URWs are local processes that hop, at each time-step, on a neighbouring node picked from a uniform probability distribution, locally maximizing the entropy rate, or dispersion, over the network. On the other hand, it is often of interest to globally maximize the spreading of a random walk over the network, such that it assumes uniform probability distribution over all paths. A process that achieves this has been proposed about a decade ago~\cite{Gomez-Gardenes2008,Burda2009}, named maximum entropy random walk (MERW), and so far used in a myriad of applications: testing network robustness~\cite{Demetrius2005,Delvenne2011} and navigability~\cite{Lin2014,Battiston2016,Estrada2021,Wang2021}, detecting connectivities and communities~\cite{Ochab2012,Ochab2012a,Xu2016,Adam2019}, benchmarking optimal mass transportation strategies~\cite{Chen2017}, predicting disease associations~\cite{Niu2018}, assessing molecular trapping~\cite{Peng2014} and neutral quasispecies evolution in biology~\cite{Smerlak2021} to name just a few. Such \deletetext{a}MERW, although characterized by local movements, i.e., it jumps to neighbouring nodes, it is non-local in the choice of the next visited node. Indeed, it needs to know the topology of the entire network to globally maximize the entropy rate~\cite{Gomez-Gardenes2008,Burda2009,Sinatra2011}.
Although MERWs are extremely useful to test spreading properties of other processes~\cite{Gomez-Gardenes2008,Sinatra2011}, they require, as mentioned, global knowledge of the embedding network. For this reason, several attempts have been made to construct optimal random walks with only local information~\cite{Sinatra2011} obtaining \textit{almost} maximum entropy random walks with step probabilities proportional to a power of the degree of the target node. }%
\newtext{A fundamental property of random walks is their ability to \textit{homogeneously} spread over the whole network. Mathematically, the spreading capability of a random walk can be characterized by measuring the entropy production rate. In many scenarios, it is indeed of uttermost importance to design random walks that maximise such entropy production rate in order to spread the most homogeneously. To picture this, imagine to have equal-size groups of random walks with different colors running on a network; at each time, the most homogeneous spreading is obtained with an equal proportion of colors on every node. Such a well mixing, or maximal dispersion, turns out to be particularly useful when information about a node state (e.g., its healthy or infected condition, its availability, etc.) needs to be homogeneously spread to all other nodes in the network~\cite{Colizza2007, Gomez-Gardenes2008,Sinatra2011}, a sought-after property for transportation and ad-hoc networks~\cite{toh2002}.

It is known that a random walk achieves maximal dispersion when it travels all trajectories of the same length with uniform probability. Such a property characterises the so-called maximum entropy random walk (MERW)~\cite{Gomez-Gardenes2008,Burda2009} and is used in a myriad of practical cases: testing network robustness~\cite{Demetrius2005,Delvenne2011} and navigability~\cite{Lin2014,Battiston2016,Estrada2021,Wang2021}, predicting links and communities~\cite{Ochab2012,Ochab2012a,Xu2016,Adam2019} as well as disease associations~\cite{Niu2018}, or assessing neutral quasispecies evolution in biology~\cite{Smerlak2021}, to name a few. However, in order to build \deletetext{a}MERW, it is required to know the topology of the whole network before even exploring it~\cite{Gomez-Gardenes2008,Burda2009,Sinatra2011}. Apart from the expensive computational cost of defining the stepping rule of \deletetext{the}MERW (based on calculating dominant eigenvalue and eigenvector of a $N \times N$ matrix, with $N$ the number of nodes), having global knowledge of the network beforehand is an heavy drawback that makes \deletetext{the}MERW totally inadequate on networks whose structure cannot be entirely determined a priori or changes over time (e.g., growing networks~\cite{albert2002statistical} or temporal networks~\cite{Holme2012,Masuda2016}). It is therefore fundamental to optimally design dispersive random walks that just make use of local information while exploring the network. Several attempts have been made in this direction, although so far finding only \emph{approximate} solutions (see for example~\cite{Sinatra2011}).}

%Our result
In this Letter, \newtext{we solve this longstanding problem} by proposing \deletetext{we propose}an adaptive random walk (ARW) that \textit{locally} updates its stepping rule based on the structure of the explored network. Without requiring any prior knowledge of the whole topology, \deletetext{the}ARW outperforms \deletetext{the}MERW, as it is maximally dispersive on every portion of the network visited and not only on the whole graph. Via a bridge between large deviation theory and network science, \deletetext{a}MERW can be seen as a rare event of \deletetext{an}URW. We \deletetext{hence} exploit this to construct \deletetext{the}ARW as a single-trajectory rare-event sampling algorithm~\cite{Coghi2021,Coghi2022} that only makes use of local information available to adapt itself---changing its transition probabilities step by step---in order to best spread on the network. \newtext{Shaping the random walk by only gathering local information is an outstanding property that makes ARW the only sensible choice for optimising the spreading on networks with time-varying topology and on heterogeneous networks. Indeed, imagine to have a network composed by two main modules, connected to each other and topologically very different. \deletetext{A}MERW initialised on one of the modules is not able to maximise dispersion while exploring it as its stepping rules are based on the `averaged' structure of the whole network. On the contrary, ARW is optimally dispersive in every exploration phase.} \newtext{In the following, this is made evident by showing that}\deletetext{ We show that, while the random walks move and discover the underlying network,} \deletetext{the}ARW has an entropy production rate closer to the maximal one on the visited portion of the graph if compared with \deletetext{an}URW and \deletetext{a}MERW. Moreover, when the network is fully explored, \deletetext{the}ARW and \deletetext{the}MERW have similar mixing properties and, in the thermodynamic limit, they become the same\deletetext{ process}.

%%Bridging large deviation theory and network science

%Bridge Observable and Entropy production rate

We start by considering an URW $X = (X_1$, $X_2$, $\cdots$, $X_n)$ on a finite connected and undirected graph $G=(V,E)$ characterized by a set of \deletetext{vertices }\newtext{nodes} $V$ and a set of \newtext{links} \deletetext{edges }$E$. The topology of the graph is encoded in the adjacency matrix 
$A=\left\lbrace a_{ij} \right\rbrace$, where $a_{ij} = 1$ if the nodes $i$ and $j$ are connected, and $a_{ij}=0$ otherwise. \newtext{We also define the degree of node $i$ as $k_i = \sum_{j \in V}a_{ij}$.}
\deletetext{The}URW dynamics is determined by the stochastic transition matrix $\Pi_{\newtext{U}}$, with components
\begin{equation}
\label{eq:URW}
\left( \pi_{\newtext{U}} \right)_{ij} = \frac{a_{ij}}{k_i} \ ,
\end{equation}
\deletetext{where $k_i = \sum_{j \in V}a_{ij}$ is the degree of node $i$,} describing the probability of \deletetext{the}URW to move from $X_\ell = i$ at time $\ell$ to $X_{\ell+1} = j$ at time $\ell+1$. We focus on \newtext{a particular} \deletetext{the }dynamical observable \newtext{that characterises the entropic content of a random walk trajectory, }
\begin{equation}
\label{eq:Obs}
C_n = \frac{1}{n} \sum_{\ell = 1}^n \ln k_{X_\ell} \ .
\end{equation}
\deletetext{which is purely additive and, as such, only depends on the visited nodes at each time step.} %\newtext{Such an observable characterises the information content of an URW trajectory.}
\newtext{Indeed, apart from boundary terms that do not influence our discussion, $C_n$ is the logarithm of the probability of \deletetext{the}URW trajectory divided by the number of time steps $n$. Noticeably, by}
\deletetext{
Because of the ergodicity of \deletetext{the}URW, guaranteed by the properties of the graph $G$, $C_n$ converges with probability $1$ to the ergodic average
\begin{equation}
\label{eq:Ave}
\left\langle \ln k \right\rangle_{\rho} = \sum_{i \in V} \rho_i \ln k_i \coloneqq c^* \ ,
\end{equation}
where $\rho = \{ \rho_i \} $ is the stationary distribution of \deletetext{the}URW $X$, in the long-time limit $n \rightarrow \infty$. }%
\deletetext{The observable $C_n$ in Eq.\ \eqref{eq:Obs} plays a pivotal role in this Letter. Indeed, up to boundary terms that do not influence our discussion and thanks to Eq.~\eqref{eq:URW}, we can rewrite $C_n = - \ln( \mathbb{P}(X_1,X_2,\cdots,X_n))/n$, i.e., as a function of the probability of the trajectory of 
\deletetext{an}URW. }%
\deletetext{T}\newtext{t}aking the long-time limit of the average over all paths of $C_n$\newtext{,} we get the so-called Kolmogorov--Sinai entropy production rate $h_{\newtext{U}}$~\cite{Cover66,Gomez-Gardenes2008,Sinatra2011}, i.e., \newtext{$h_U = \lim_{n \rightarrow \infty} \left\langle C_n \right\rangle$}\newtexttwo{, interpreted as }\newtext{the mean information generated per time step}\newtexttwo{. For a generic ergodic random walk, it can be written as}
\deletetext{
\begin{equation}
\label{eq:KSEntr}
h = \lim_{n \rightarrow \infty} \left\langle C_n \right\rangle \ ,
\end{equation} 
}
\deletetext{which, for an URW, takes the form in Eq.\ \eqref{eq:Ave}, and can be interpreted as the mean information per time step generated by the random walk.}
\deletetext{Furthermore, for general ergodic random walks, the form in Eq.\ \eqref{eq:KSEntr} is known~\cite{Cover66}, with abuse of notation, to be given by}
\begin{equation}
\label{eq:CoverKSEntr}
h = - \sum_{i,j} \rho_i \pi_{ij} \ln \pi_{ij} \ ,
\end{equation}
\newtext{where $\rho = \{ \rho_i \} $ and $\pi=\{ \pi_{ij} \}$ are the stationary distribution and the transition probability matrix of}\newtexttwo{ the random walk.} \deletetext{It is then clear that}\newtext{Eventually,} the observable $C_n$ is a random variable of the random walk process that represents the fluctuating version of $h_{\newtext{U}}$, viz.\ the fluctuating trajectory entropy~\cite{Seifert2005,Touchette2009}.

%Large deviations of URWs

The finite-time fluctuating nature of $C_n$ around \deletetext{$c^*$ }\newtext{its typical value $h_U = \sum_{i \in V} \rho_i \ln k_i$} is \newtext{of interest} \deletetext{what interest us} here. A complete understanding of the fluctuations is given by the probability density $P_n(c) \coloneqq P(C_n = c)$, which is known to have the large deviation form
\begin{equation}
\label{eq:Rate}
P_n(c) = e^{-n I(c) + o(n)} \ ,
\end{equation}
\newtext{with the non-negative large deviation rate function $I(c)$ characterizing the leading behavior of $P_n(c)$ and }\deletetext{where }$o(n)$ denot\deletetext{es}\newtext{ing} corrections smaller than linear in $n$. The focus thus moves onto studying \deletetext{the non-negative large deviation rate function }$I$\deletetext{ given by the leading exponential behaviour} in Eq.\ \eqref{eq:Rate}, \deletetext{characterized by having}\newtext{which has} a unique zero \deletetext{in $c^*$}\newtexttwo{at }\newtexttwo{$c^*= $} $h_U$. The rate function can be calculated by means of the so-called G\"{a}rtner--Ellis theorem~\cite{DenHollander2000,Touchette2009,Dembo2010}, which states that $I$ is given by the Legendre--Fenchel transform of the scaled cumulant generating function (SCGF)
\begin{equation}
\label{eq:SCGF}
\Psi(s) = \lim_{n \rightarrow \infty} \frac{1}{n} \ln \mathbb{E} \left[ e^{n s C_n} \right] \ ,
\end{equation}
as this last is differentiable for a finite graph~\cite{Dembo2010}. In particular, as \deletetext{the}URW is an ergodic Markov process, the SCGF can be obtained as
\begin{equation}
\label{eq:SpectralSCGF}
\Psi(s) = \ln \zeta_s ,
\end{equation}
where $\zeta_s$ is the dominant eigenvalue of the so-called tilted matrix $\tilde{\Pi}_s = \left\lbrace \left( \tilde{\pi}_s \right)_{ij} \right\rbrace$, with components
\begin{equation}
\label{eq:TiltedMatrix}
\left( \tilde{\pi}_s \right)_{ij} = \pi_{ij} e^{s \ln k_i} = \pi_{ij} k_i^s = a_{ij} k_i^{s-1} \ .
\end{equation}
Hence, the likelihood of fluctuations can be studied using the SCGF $\Psi$ rather than the rate function $I$.

However, calculating the probability of fluctuations is only a first step towards the prediction and control of rare events. It is indeed important to also understand how these extreme events are created in time. In this context, we construct the driven process~\cite{Chetrite2013,Chetrite2015,Chetrite2015a} associated with a given fluctuation $C_n = c$\newtext{. This process} \deletetext{that here }is a locally-biased version of \deletetext{the}URW~\cite{Chetrite2015,Coghi2019} \deletetext{whose} \newtext{and its} transition probability matrix is given by
\begin{equation}
\label{eq:DrivenProcess}
\left( \pi_s \right)_{ij} = \frac{\left( \tilde{\pi}_s \right)_{ij} r_s(j)}{r_s(i) \zeta_s} = \frac{a_{ij} k_i^{s-1} r_s(j)}{r_s(i) \zeta_s} \ ,
\end{equation}
where $r_s$ is the right eigenvector associated with $\zeta_s$. The driven process is still Markovian and ergodic and can be interpreted as the effective dynamics of the subset of paths of \deletetext{the}URW leading to a fluctuation $C_n = c$~\cite{Chetrite2015,Coghi2019,Gutierrez2021}; to match such a fluctuation~\cite{Touchette2005}, the Laplace parameter $s$ must satisfy
\begin{equation}
c=\Psi'(s) \ .
\label{eq:Duality}
\end{equation}
Eventually, the entropy rate of the driven process can be obtained taking Eq.\ \eqref{eq:DrivenProcess} and plugging it into Eq.\ \eqref{eq:CoverKSEntr} and can be expressed in terms of the SCGF~\cite{Coghi2019} as 
\begin{equation}
\label{eq:EntrRateSCGF}
h(s) = \Psi(s) + (1-s)\Psi'(s) \ .
\end{equation}
\newtext{In }\newtexttwo{Sect.\ II of }\newtext{the Supplemental Material (SM) we show that $h(s)$ in Eq.\ \eqref{eq:EntrRateSCGF}} \deletetext{This expression can be analytically proved to have }\newtext{has} a global maximum for $s=1$, i.e., 
\begin{equation}
\label{eq:MaxEntrRate}
h(1) = \Psi(1) = \ln \zeta_1 \ ,
\end{equation}
\deletetext{shown in Fig.\ \ref{fig:SCGF},} where $\zeta_1$ is the dominant eigenvalue of the adjacency matrix $A$. 
Replacing $s=1$ in the driven process, Eq.\ \eqref{eq:DrivenProcess} gives \deletetext{the}MERW
\begin{equation}
\label{eq:MERW}
\left( \pi_1 \right)_{ij} = \frac{ a_{ij} r_1(j)}{r_1(i) \zeta_1} \ ,
\end{equation}
allowing us to interpret \deletetext{a}MERW on a network as a biased random walk creating a \deletetext{particular }rare event fluctuation---given by replacing $s=1$ in Eq.\ \eqref{eq:Duality}---of \deletetext{an}URW~\cite{Coghi2019}.

%% The ARW

% Sampling the fluctuation

This result shows, on the one hand, that we can sample a particular rare event of \deletetext{an}URW by simulating \deletetext{a}MERW and, on the other hand, that \deletetext{a}MERW can in principle be obtained from \deletetext{an}URW by opportunely conditioning on a certain rare event of the observable $C_n$ in Eq.\ \eqref{eq:Obs}. The latter observation is key to introduce our adaptive random walk (ARW)\deletetext{,}. \deletetext{which through successive local adaptations of \deletetext{an}URW converges to \deletetext{a}MERW on the whole network.}\deletetext{In fact, a}As we will show in the following, such ARW\newtext{, through successive local adaptations of URW,} reaches maximum entropy while exploring the network, i.e., much before the entire graph has been visited\deletetext{.}\newtext{, and eventually converges to MERW on the whole network.} To construct \deletetext{the}ARW, we develop here an algorithm based on a rare-event sampling scheme~\cite{Borkar2003,Ahamed2006,Ferre2018,Ferre2018a,Coghi2021,Coghi2022}. According to this, the random walk updates its transition probability matrix at each time step, in order to typically visit a specific rare event of \deletetext{the}URW, obtained fixing $s=1$ in Eq.\ \eqref{eq:Duality}. We refer the reader to \newtexttwo{Sect.\ I of }\deletetext{Supplemental Material }\deletetext{(}the SM\deletetext{)} for a general formulation of the sampling algorithm valid for all additive observables and $s \in \mathbb{R}$.

% Minimal technical details on the adaptive scheme

Formally, \deletetext{the}ARW is a discrete-time process $Y=(Y_1$, $Y_2$, $\cdots$, $Y_n)$, where $Y_n \in V$ is the position of \deletetext{the}ARW on the graph at time $n$. The core of \deletetext{the}ARW is based on an adaptive power method to solve the following dominant eigenvalue equation
\begin{equation}
\label{eq:EigProb}
\tilde{\Pi}_1 r_1 = \zeta_1 r_1 \ , 
\end{equation}
which is known to be cardinal to construct \deletetext{the}MERW in Eq.~\eqref{eq:MERW}. 
%In fact, in a large deviation context, one seeks to solve Eq.\ \eqref{eq:EigProb} to determine the SCGF $\Psi(1)$ via Eq.\ \eqref{eq:SpectralSCGF}. As previously discussed, this SCGF is linked to the probability of the rare event $c=\Psi'(1)$, typically generated by a MERW.
More in detail, our adaptive power method simulates single Markov chain transitions with importance sampling~\cite{Bucklew2004}. In particular, supposing that \deletetext{the}ARW is located on the node $i$ at time $n$, i.e., $Y_n = i$, the next step is proposed according to an estimate of \deletetext{the}MERW in Eq.~\eqref{eq:MERW} that reads
\begin{equation}
\label{eq:EstimateDriven}
\left( \pi_1^{(n)} \right)_{ij} = \frac{1}{Z} \frac{a_{ij} r_1^{(n)}(j)}{r_1^{(n)}(i) r_1^{(n)}(i_0)} = \frac{a_{ij} r_1^{(n)}(j)}{\sum_{j' \in V} a_{ij'} r_1^{(n)}(j')} \ ,
\end{equation}
where $Z$ is the normalisation factor, $i_0$ is an a-priori fixed node, and $r_1^{(n)}$ is the $n$-th time estimate of the eigenvector centrality~\cite{Coghi2021,Coghi2022}. This last is given by the stochastic-approximation~\cite{Borkar1998} formula
\begin{equation}
\label{eq:StochApproxRight}
\begin{split}
&r_1^{(n+1)}(i) = r_1^{(n)}(i) \,+ \\ 
&\hspace{0.6cm} + \newtexttwo{\lambda}\deletetext{a}(n) \mathbf{1}_{Y_n=i} \left( \frac{\sum_{j \in V} a_{ij} r_1^{(n)}(j)}{r_1^{(n)}(i_0)} - r_1^{(n)}(i) \right)
\end{split}
\end{equation}
based \deletetext{on the learning rate $a(n)$ and on an asynchronous update via the indicator function $\mathbf{1}_{Y_n=i}$}\newtext{on an asynchronous update via the indicator function $\mathbf{1}_{Y_n=i}$ and on the learning rate $\newtexttwo{\lambda}\deletetext{a}(n)$}. \newtext{The indicator function selects}\deletetext{In other words,} the $i$-th component of the eigenvector centrality \deletetext{is}\newtext{to be} updated only when the process $Y$, at the $n$-th time step, has visited node $i$. \newtext{Additionally, the learning rate $\newtexttwo{\lambda}\deletetext{a}(n)$---commonly used in stochastic approximation protocols~\cite{Borkar1998,Benaim1999,Borkar2003,Ahamed2006}---expresses how much of the information that has been learnt up to time $n$ is used to update $r_1$ in the next time step.} \deletetext{Also n}\newtext{N}ote that, in the following, all ARW simulations are obtained by 
\newtexttwo{using} 
a learning rate $\newtexttwo{\lambda}\deletetext{a}(n)=1/((n+1)^{\beta})$, with $\beta=0.1$, in Eq.~\eqref{eq:StochApproxRight}. 
\newtexttwo{Although there is no theory to a-priori determine the learning rate, 
there are mathematical conditions that $\lambda$ needs to satisfy, and one can carry out numerical simulations on benchmark networks to finely tune the value of $\beta$. We show how to do so in Sections III and VI of the SM (see Fig.\ S2-S6, Fig.\ S9, and Table S1), further noticing 
that our approach, in order to set off an `optimal' value of $\beta$, does not require to compare the performance of ARW with MERW, or with any other spreading process that requires global knowledge of the network.  %optimal spreading processes on the networks
}

% Comparison between ARW and MERW

\begin{figure*}[t]
\centering
\includegraphics[width=\textwidth]{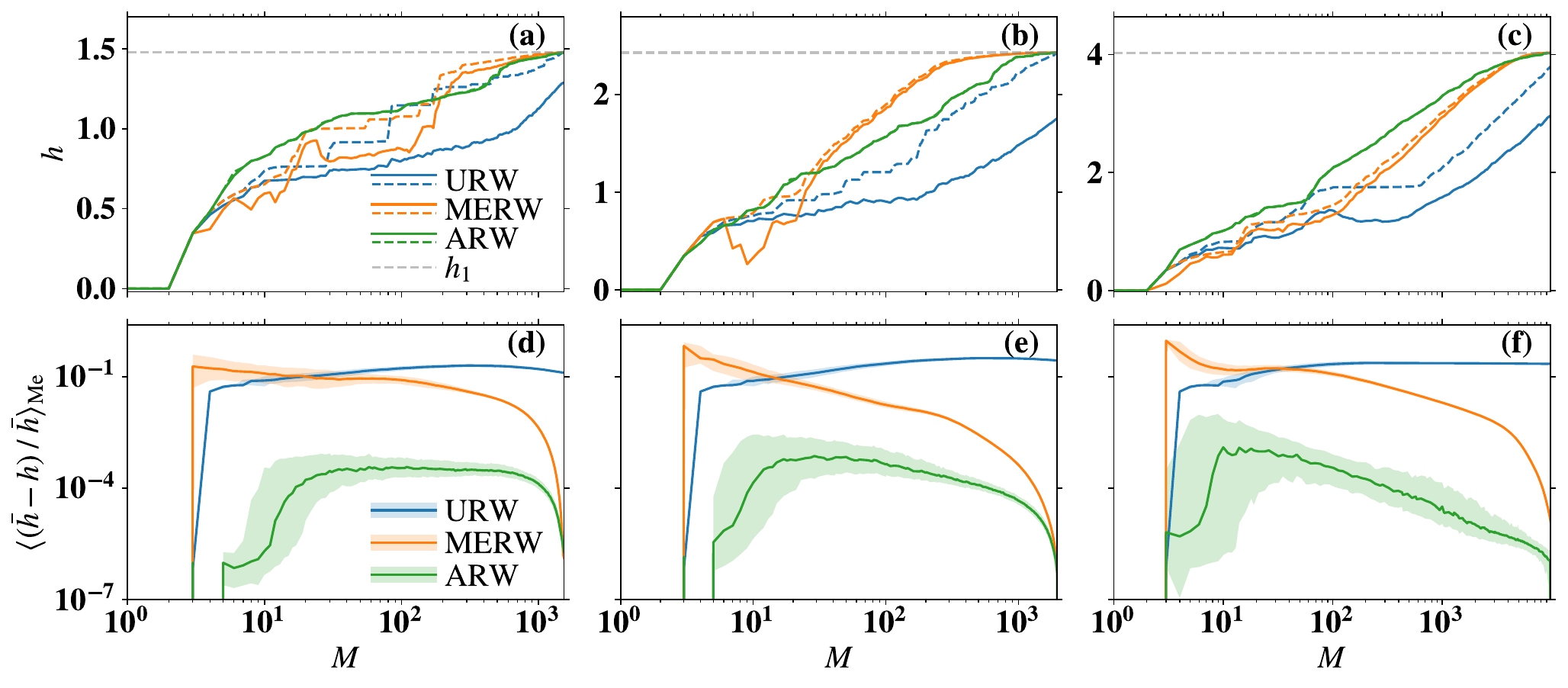}
\caption{(a-c)
The entropy production rate $h$, calculated as in Eq.~\eqref{eq:entrratevisited} of single trajectories of ARW, MERW and URW (solid lines) is compared to the corresponding optimal entropy production rate $\bar{h}$ on the discovered graph (dashed lines).
(d-f) Median (solid lines) and first and third quartiles (shaded area) of the normalized differences $(\bar{h}-h)/\bar{h}$ over an ensemble of $1000$ trajectories of ARW, MERW, and URW. 
Results are shown for random walks running on the giant connected component of an Erd\"{o}s--R\'{e}nyi random graph with $1000$ nodes and average degree $3$ in (a,d), a Barabasi--Albert with $1000$ nodes and $m=\newtext{2}$\deletetext{4}~\cite{albert2002statistical} in (b,e), and an air transportation network~\cite{guimera2005worldwide} with $3618$ nodes and $14142$ links in (c,f).}
\label{fig:ARWMERWURW}
\end{figure*} 

\deletetext{The}ARW is randomly initialized on a node of the network with a normalized random right eigenvector $r_1^{(0)}$ and evolves according to the fully local rules in Eq.\ \eqref{eq:EstimateDriven} and Eq.\ \eqref{eq:StochApproxRight}. At each time step it tends to optimize the spreading---aiming at\deletetext{the} maximum entropy production---on the portion of the network visited. In the long-time limit, it will eventually converge to \deletetext{the}MERW of Eq.\ \eqref{eq:MERW} as $r_1^{(n)} \rightarrow r_1$, $r_1^{(n)}(i_0) \rightarrow \zeta_1$, and $Z \rightarrow 1$ . We insist on the fact that differently from \deletetext{the}MERW, \deletetext{the}ARW does not need to know the full topology of the network since the beginning, as it learns it on the run. Thanks to this, its entropy production rate stays always close to the maximum rate on the visited portion of the graph. This is drastically different from \deletetext{a}MERW which does not maximize the entropy while it explores the graph, but reaches optimal spreading only when the whole network has been visited.

We show this in Fig.\ \ref{fig:ARWMERWURW}(a)--(c) where we compare the spreading properties of single trajectories of ARW, MERW, and URW on two network models, namely  Erd\"{o}s--R\'{e}nyi, and Barabasi--Albert, and on a real-world man-made network. The last network describes an air transportation system: each node is a city and two nodes are connected if at least one airplane flew between the two cities in the time window \deletetext{[11/01/2000 - 10/01/2001]}\newtexttwo{[11 Jan 2000 - 10 Jan 2001]}~\cite{guimera2005worldwide}. \newtext{In this last context, maximizing entropy production rate allows manufacturers, for example, to homogeneously spread goods around their factories.} Each of the three processes is initialized on a randomly selected node of the network, and evolves according to its transition probability matrix. As the walker moves \newtext{hopping through}\deletetext{to} previously-unvisited \deletetext{nodes }\newtext{links}, we calculate its entropy production rate $h$ (solid line) and compare it with the optimal $\bar{h}$ (dashed line) given by the logarithm of the dominant eigenvalue of the adjacency matrix associated with the portion of graph made by all (and only) previously visited links. The entropy production rate $h$ of each process is calculated via a modified version of Eq.\ \eqref{eq:CoverKSEntr}, that is
\begin{equation}
\label{eq:entrratevisited}
h(M) = - \sum_{(i,j) \in E(n)} \rho_i(M) \pi_{ij}(M) \ln \pi_{ij}(M) \ .
\end{equation}
This takes into account the number of visited links $M = |E(n)|$ up to time $n$, where $E(n)$ is the set of visited links, $\pi_{ij}(M) = \pi_{ij}/ ( \sum_{(i,j') \in E(n)} \pi_{ij'})$ if the link $(i,j)$ is in $E(n)$, \deletetext{and }\newtext{while} $0$ otherwise, and the stationary distribution $\rho(M)$ is calculated as the left eigenvector of $\Pi(M) = \left\lbrace \pi_{ij}(M) \right\rbrace$. As the solid \newtext{green} line is \newtext{indistinguishable from}\deletetext{ closer to} the corresponding dashed one\deletetext{ for the ARW }\deletetext{with respect to the MERW and the URW, we notice that}\newtext{,} \deletetext{the}ARW \newtext{has always}\deletetext{ reaches earlier}---while exploring the network---\newtext{optimal}\deletetext{ best} spreading performances.
\newtext{On the contrary, by comparing the corresponding solid and dashed lines, the performances of MERW and URW are always sub-optimal. In particular, MERW reaches maximum entropy production---comparable to that of ARW---only}
\deletetext{W}\newtext{w}hen the whole graph has been visited%
\deletetext{, the ARW performance is comparable to that of the MERW with the entropy production rate in Eq.\ \eqref{eq:MaxEntrRate}, whilst the URW converges to Eq.\ \eqref{eq:CoverKSEntr}}.

Moreover, in Fig.\ \ref{fig:ARWMERWURW}(d)--(f) we plot the median (solid line) of the relative difference between the entropy production rate $h$ and the optimal $\bar{h}$, together with first and third quartiles (shaded area). The median and the quartiles are calculated over an ensemble of $1000$ trajectories and are used in place of the mean and standard deviation because of the unknown distribution of $h$ around $\bar{h}$. This gives further evidence of the fact that \deletetext{the}ARW performs better than \deletetext{the}MERW (and \deletetext{the}URW) at maximizing the spreading while discovering the structure of the graph. Fig.\ \newtexttwo{S7}\deletetext{2} of the SM shows the entropy production rates of ARW, MERW, and URW on the graph induced by the visited nodes, i.e, including links that have not yet been visited. In this case, \deletetext{the}ARW is still outperforming MERW and URW on the Erd\"{o}s--R\'{e}nyi graph, but is only marginally better on the \deletetext{scale-free and the real-world network }\newtext{other networks}.

%Mechanism behind spreading

\begin{figure}[b]
\centering
\includegraphics[width=1.0\columnwidth]{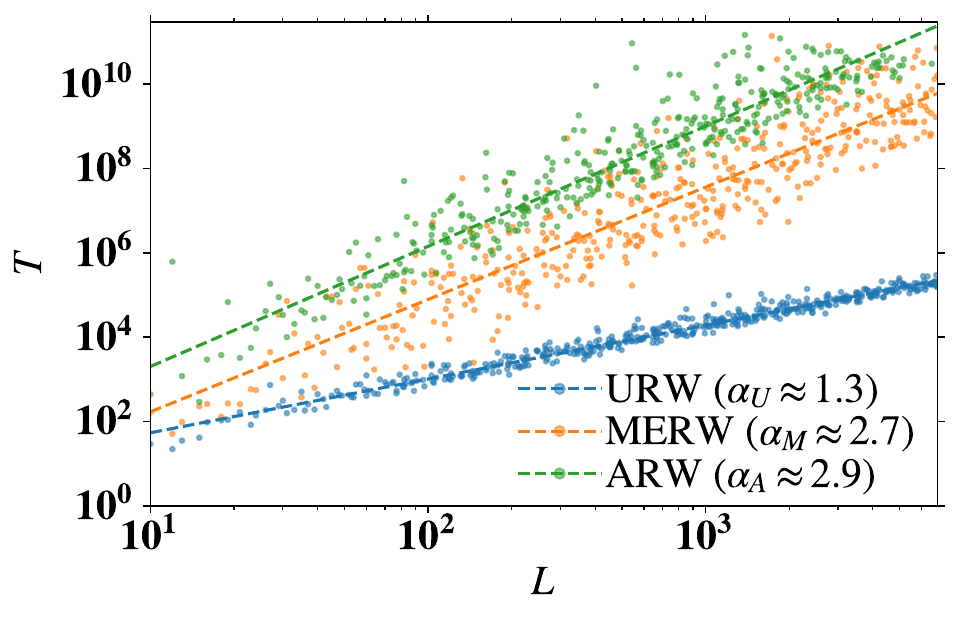}
\caption{Coverage time $T$ for ARW, MERW, and URW as a function of $L$, i.e., the number of links in the giant connected component of Erd\"{o}s-R\'{e}nyi graphs with average degree $3$ and increasing size.}
\label{fig:CovTime}
\end{figure}

The optimal dispersion of the fully-local ARW on the visited links of the network comes at a price: \deletetext{the}ARW takes longer than URW and MERW to cover the whole graph. This is consequence of (i) an initial so-called warm-up phase in which \deletetext{the}ARW is localised in the region where it \deletetext{has been}\newtext{was} initialised, and (ii) a typical exploration time of the network. During the warm-up phase, our process finely tunes the eigenvector centrality and the transition probability matrix to set off an efficient exploration of the network. As shown in Fig.\ \newtexttwo{S8}\deletetext{3} of the SM, where we plot the average number of time steps needed to discover new links, \deletetext{the}ARW remains indeed localised in the first few visited links\deletetext{ of the network}. This is also the main reason why the network coverage time---the number of time steps to visit all links---is, on average, $10^2-10^4$ time steps longer for \deletetext{an}ARW than for \deletetext{a}MERW or \deletetext{an}URW (see Fig.\ \ref{fig:CovTime}).
We find that the coverage time $T$ is related to the total number of links $L$ \deletetext{by the relation }\newtext{as} 
\begin{equation}
\label{eq:CharTime}
T \propto L^\alpha \ ,
\end{equation}
where $\alpha$ is the scaling exponent. Remarkably, after the initial warm up---which is evident by the overall upward shift of \deletetext{the}ARW power-law fit---the scaling exponents of ARW and MERW are similar ($\alpha_A \approx 2.9$, $\alpha_M \approx 2.7$), but larger than the one of \deletetext{the}URW which has no global constraint to satisfy ($\alpha_U \approx 1.3$). \newtext{We remark that optimising entropy production rate and coverage time are two different tasks with the former much harder than the latter. Although having both properties is certainly appealing, our main goal is to optimise dispersion. }\newtext{However, in an attempt to also optimize coverage time, }\deletetext{As a last remark,}we point out that in our studies\deletetext{of the ARW}, the initial right eigenvector $r_1^{(0)}$ plays a key role in determining the initial warm-up time and the overall accuracy of \deletetext{the}ARW in optimizing the spreading while exploring the network (see Fig.\ \newtexttwo{S10}\deletetext{4} of the SM). We hope that our work will stimulate further investigations to explore the trade-off between optimal spreading accuracy and exploration times.

%we collect several plots that compare the spreading and coverage time performances of another adaptive random walk, dubbed $\overline{\text{ARW}}$, with MERW, URW, and the ARW discussed so far. The $\overline{\text{ARW}}$ evolves in time according to the same algorithm of the ARW, but differently from the latter it is initialized with a non-normalized right eigenvector. Noticeably, the warm-up time---therefore, also the coverage time---is greatly reduced with respect to the ARW at the expense of a spreading that is comparable with a MERW and hence not optimal while exploring the network.

%%Conclusion

In this Letter, we have proposed an adaptive random walk that has optimal spreading properties, outperforming the well-known MERW. Via a large-deviation tilting on the fluctuating trajectory entropy observable, \deletetext{the}ARW typically observes a maximum entropy production rate while exploring the network, exploiting only local information. 
% The ARW is instead based on a random walk that by means of only local rules adapts itself to be a maximum entropy random walk while visiting the network, without requiring any prior knowledge of it. The adaptive step is the fundamental ingredient of the process. Via a large-deviation tilting on the fluctuating trajectory entropy observable, the adaptive step allows the ARW to typically observe a maximum entropy production rate, which qualifies as a rare event of the URW. 
% We have also shown two key features of the ARW dynamics: the initial warm-up time---independent on the size of the graph---that allows the ARW to set off an optimal spreading on the network, and the scaling exponent of the network coverage time, which is similar to that of the MERW. 
Besides the theoretical novelty driven by a large deviation study of random-walk rare events, we believe that \newtext{our work can be a fundamental step towards the study of}\deletetext{ the ARW might find useful applications to} network information spreading~\cite{Gao2016,Peixoto2019} in all such cases where no prior knowledge on the network is available, or when the network is changing in time~\cite{Holme2012,Masuda2016,Holme2019}.
\deletetext{It }\newtext{ARW} could also be used to study dispersion properties of other dynamical processes on real networks, e.g., aiming at optimal exploration in congested networks~\cite{Manfredi2018,Carletti2020}. Furthermore, the algorithm at the core of \deletetext{the}ARW could also be used to sample other rare-event fluctuations associated with any additive observables of random walks~\cite{Coghi2021,Coghi2022}.

\section{Code}%\deletetext{s}}

All the code\deletetext{s} used in the manuscript \deletetext{are }\newtext{is} available at \url{https://github.com/gabriele-di-bona/ARW}.

\newtext{
\section{Acknowledgments}

FC is deeply grateful to Pierpaolo Vivo and Hugo Touchette for valuable comments and suggestions in the writing stage of the manuscript.
}

\clearpage

\onecolumngrid

%\chapter{Supplemental Material}

\section{\large{Supplemental Material}}

\vspace{1cm}

\section{Adaptive random walk algorithm}
\label{sec:ARW}

In this Section we give details on the adaptive random walk process $Y=(Y_\ell)_{\ell=1}^n$ which is based on an efficient algorithm to sample large-deviation rare events~\cite{Coghi2021,Coghi2022}. Differently from the main text, we will not restrict to the particular case of a MERW, but we will describe the numerical scheme in its full generality. However, all the results in the main text (MT) can be obtained fixing $f(X_\ell)=k_{X_\ell}$ in Eq.~\eqref{eq:Obs} and replacing $s=1$ in Eq.~\eqref{eq:TiltMatr} and following.

We consider a Markov chain $X = (X_\ell)_{\ell=1}^n$ evolving in a finite discrete state space $\Gamma$ of $N$ states (this is the graph $G=(V,E)$ in the MT) according to the (irreducible and aperiodic) transition matrix $\Pi$ with elements $\pi_{ij}$ that characterize the probability of going from a state $X_\ell = i$ at time $\ell$ to a state $X_{\ell+1} = j$ at time $\ell+1$. Associated with the Markov chain $X$, we consider a purely time-additive observable of the general form
\begin{equation}
\label{eq:Obs}
C_n = \frac{1}{n} \sum_{\ell=1}^n f(X_\ell) \ ,
\end{equation}
where $f$ is any function of the state. As mentioned in the MT, one can estimate the likelihood of fluctuations of $C_n$ by calculating the scaled cumulant generating function (SCGF) in Eq.\ \deletetext{(7)}\newtext{(5)}MT (and Eq.\ \deletetext{(8)}\newtext{(6)}MT) and by Legendre--Fenchel transforming it to obtain the large deviation rate function, i.e., the leading exponential behaviour of the probability density in Eq.\ \deletetext{(6)}\newtext{(4)}MT. However, this procedure is known to be difficult since it involves the calculation of the dominant (Perron--Frobenius) eigenvalue $\zeta_s$ of the non-negative tilted matrix $\tilde{\Pi}_s = \left\lbrace \left( \tilde{\pi}_s \right)_{ij} \right\rbrace$ having elements 
\begin{equation}
\label{eq:TiltMatr}
\left( \tilde{\pi}_s \right)_{ij} = \pi_{ij} e^{s f(i)} \ ,
\end{equation}
see \cite{Touchette2009,Dembo2010,Chetrite2015,Coghi2021} for further details. This problem can be mathematically formalized as finding the solution of the following spectral equation
\begin{equation}
\label{eq:EigenProblem}
\tilde{\Pi}_s r_s = \zeta_s r_s \ ,
\end{equation}
where along with $\zeta_s$ we have the right eigenvector $r_s$. Both these are fundamental to define the driven process, which is characterized by a transition matrix whose components are
\begin{equation}
\label{eq:DrivenTransition}
\left( \pi_s \right)_{ij} = \frac{\left(\tilde{\pi}_s \right)_{ij} r_s(j)}{r_s(i) \zeta_s} = \frac{\pi_{ij} e^{s f(i)} r_s(j)}{r_s(i) \zeta_s} \ ,
\end{equation}
as in the first equality of Eq.\ \deletetext{(10)}\newtext{(8)}MT.

In the following, we show a numerical method that approximates the dominant right eigenvector $r_s$ and, consequently, the dominant eigenvalue $\zeta_s$, and so the SCGF $\Psi$ too (Eq.\ \deletetext{(8)}\newtext{(6)}MT), by making use of an adaptive stochastic power-method scheme. The core of the algorithm is rooted on works of reinforcement learning for risk-sensitive control of Markov chains \cite{Borkar2002,Borkar2003,Ahamed2006,Basu2008} and was also recently adapted to estimate large deviation functions of continuous-time diffusion processes \cite{Ferre2018,Ferre2018a}. Here, we repropose a discussion of the method that appeared in \cite{Coghi2021}.

A bare power method scheme recursively multiplies an arbitrary vector $v_0 \coloneqq r_s^{(0)}$ (the superscript $(0)$ refers to the zero-th step in the numerical scheme) by the transition matrix $\tilde{\Pi}_s$. In this case, we can write 
\begin{equation}
\label{eq:RightEigvApproxPow}
\zeta_s^{(n+1)} r_s^{(n+1)} = \tilde{\Pi}_s^{n+1} r_s^{(0)} = \tilde{\Pi}_s r_s^{(n)}
\end{equation}
where $r_s^{(n+1)}$ is the approximation at the $(n+1)$-th step of the dominant right eigenvector normalized by the $(n+1)$-th estimate of the dominant eigenvalue $\zeta_s^{(n+1)}$. This can be calculated as 
\begin{equation}
\label{eq:EigvApproxPow}
\zeta_s^{(n)} = r_s^{(n)}(i_0) \ ,
\end{equation}
where, a priori, the component $i_0$ is arbitrarily chosen, but here we fix it to be the maximum component of $r_s^{(n)}$ in magnitude such that the right eigenvector is normalized according to its infinity norm\footnote{The infinity norm of a vector is the maximum of the vector's absolute-value components.}. As $n \ra \infty$ we then have
\begin{equation}
\label{eq:ConvPow}
r_s^{(n)} \ra r_s \hspace{1cm} \text{and} \hspace{1cm} r_s^{(n)}(i_0) \ra \zeta_s \ .
\end{equation}

Numerically, it is known that the power method scheme in Eq.~\eqref{eq:RightEigvApproxPow} is particularly inefficient for large (and very connected) state spaces, because at each step one matrix multiplication needs to be calculated. To overcome this problem, as we will see in the following, one may consider to implement a numerical scheme that simulates single Markov chain transitions rather than keeping track of the full structure of the state space in the matrix multiplication. These transitions could be naively proposed according to the tilted matrix in Eq.~\eqref{eq:TiltMatr}. Notwithstanding this, due to the presence of the exponential factor, such an approach could lead to divergences. For this reason, one considers importance sampling \cite{Bucklew2004} and simulates the $(n+1)$-th transition step according to the $n$-th estimate of the driven process in Eq.~\eqref{eq:DrivenTransition}, that is
\begin{equation}
\label{eq:DrivenEstimate}
\left. (\pi_s^{(n)})_{i j} = \frac{\pi_{i j} e^{s f(i)} r^{(n)}_s(j)}{r^{(n)}_s(i) r^{(n)}_s(i_0)} \middle/ \left( \sum_{j'=1}^N \frac{\pi_{i j'} e^{s f(i)} r^{(n)}_s(j')}{r^{(n)}_s(i) r^{(n)}_s(i_0)} \right) \right. = \frac{\pi_{i j} r^{(n)}_s(j)}{\sum_{j'=1}^N \pi_{i j'} r^{(n)}_s(j')} \ .
\end{equation}
With a simulated transition from $i$ to $j$ the estimate of the right eigenvector in Eq.~\eqref{eq:RightEigvApproxPow} therefore reads
\begin{equation}
\label{eq:RightEigvApproxAdapt}
r^{(n+1)}_s(i) = \frac{e^{s f(i)}}{r^{(n)}_s(i_0)} r^{(n)}_s(j) \frac{\pi_{ij}}{(\pi_s)^{(n)}_{i j}} =  \frac{e^{s f(i)} \sum_{j'=1}^N \pi_{ij'} r_s^{(n)}(j')}{r_s^{(n)}(i_0)} \ ,
\end{equation}
where we normalize the right eigenvector with $r^{(n)}_s(i_0)$ ($i_0$ is, once again, the maximum component of $r_s^{(n)}$ in magnitude) and, as usual in importance sampling, we unbias the transition multiplying by the likelihood ratio\footnote{Discrete Radon--Nikodym derivative.} $\pi_{ij}/(\pi_s^{(n)})_{ij}$.
%\footnote{Notice that a transition of this adaptive process could also be made according to the unbiased transition matrix $\Pi$, multiplying by the exponential tilting factor only a posteriori. In such a case the $n$-th step estimate of the driven process would not be biased and so one would not need to unbias it multiplying by the likelihood ratio.}
Notice that it is only in the limit $n \ra \infty$ that the estimate in Eq.~\eqref{eq:DrivenEstimate} tends to the driven process of Eq.~\eqref{eq:DrivenTransition}. Indeed, the normalisation appearing in Eq.~\eqref{eq:DrivenEstimate} has a pivotal role in correctly estimating the driven process transition probabilities, and it is only in the limit $n \ra \infty$ that such normalisation converges to $1$.

Eventually, convergence of Eq.~\eqref{eq:RightEigvApproxAdapt}, as in Eq.~\eqref{eq:ConvPow}, is realized making use of a stochastic approximation scheme. Stochastic approximation methods \cite{Borkar1998,Benaim1999} are discrete-time (stochastic) recursive schemes similar, in form, to gradient descent methods. They make use of an annealing schedule, also called learning rate, to quantify, at each time step, `how much' of the new information coming from an algorithm is used to update the value of an observable of interest in the next step. In our case, this observable of interest is the dominant right-eigenvector, and the stochastic approximation scheme applied to the algorithm in Eq.~\eqref{eq:RightEigvApproxAdapt} reads  
\begin{equation}
\label{eq:RightEigvApproxAdaptLearn}
r^{(n+1)}_{s} (i) = r^{(n)}_s(i) + \deletetext{a}\newtexttwo{\lambda}(n) \mathbf{1}_{Y_n = i} \left(\frac{e^{s f(i)} \sum_{j=1}^N \pi_{ij} r_s^{(n)}(j)}{r_s^{(n)}(i_0)} - r^{(n)}_s(i) \right) \ ,
\end{equation} 
where $\mathbf{1}_{Y_n = i}$ makes the update asynchronous,
%\footnote{A synchronous approximation scheme could also be implemented if we could  available for each state of the state space.}
viz.\ it selects only the state $i$ at time step $n$, and $\deletetext{a}\newtexttwo{\lambda}$ is the learning rate, or annealing schedule, which should satisfy \newtext{certain mathematical features~\cite{Borkar1998} and is here assumed to have the general form
\begin{equation}
    \label{eq:LearningRate}
    \deletetext{a}\newtexttwo{\lambda}(n) = \frac{1}{(1+n)^{\deletetext{\alpha}\newtexttwo{\beta}}} \ .
\end{equation}}
% \begin{equation}
% \label{eq:AnnSchProp}
% \lim_{n \ra \infty} \sum_{\ell=1}^n a(\ell) = \infty \hspace{2cm} \lim_{n \ra \infty} \sum_{\ell=1}^n a^2(\ell) < \infty .
% \end{equation}
A proof of the convergence of this numerical method is given in \cite{Borkar1998}, where it is shown that the numerical scheme tracks the solution of a particular nonautonomous differential equation. \newtext{As mentioned in the MT, the learning rate is a fundamental ingredient of the stochastic approximation scheme we have introduced as it really determines how much of what has been learnt up to time $n$ is used to update the right-eigenvector $r_s$ in the next time step. Generally speaking, apart from the mathematical conditions given in~\cite{Borkar1998} which ensure the mathematical convergence of Eq.~\eqref{eq:RightEigvApproxAdaptLearn}, to our knowledge there is no general recipe to determine the `optimal' form of the learning rate $\deletetext{a}\newtexttwo{\lambda}$ one should choose to have good---fast---numerical convergence of Eq.~\eqref{eq:RightEigvApproxAdaptLearn}. Indeed, $\deletetext{a}\newtexttwo{\lambda}$ may depend on many different features related to the particular model investigated: the initial condition $r_s^{(0)}$, the local structure of the state space encoded in $\pi_{ij}$, etc. Hence, an exact functional form of $\deletetext{a}\newtexttwo{\lambda}$ may be very hard to find and is beyond the scope of our work. Generally speaking, the key rule is that $\deletetext{a}\newtexttwo{\lambda}$ should not decay neither too quickly nor too slowly in time as the random walk $Y$ needs some time to learn the local features of the network, but it should not be allowed too much time to avoid slow or inaccurate convergence.}

The remarkable advantage of this adaptive stochastic numerical scheme, based on the combination of spectral methods with important sampling, is that it allows us to estimate the SCGF $\Psi$ by simulating a single trajectory of a Markov chain that learns on the run the particular rare event of interest $c$ of Eq.~\eqref{eq:Obs}, linked to $s$ by $\Psi'(s)=c$. For instance, we simulate the ARW described in the MT with Eq.\ (8) and Eq.\ (10) by fixing $s = 1$, i.e., Eq.\ \deletetext{(16)}\newtext{(14)}MT and Eq.\ \deletetext{(17)}\newtext{(15)}MT, to adaptively obtain a random walk with maximum entropy production on every visited portion of the network. Moreover, in a large deviation context, one can estimate the full SCGF by discretizing with a fine mesh the tilting parameter space, i.e., $[\dots, s - \Delta s, s, s + \Delta s, \dots]$, and running the algorithm for every point in the grid. Eventually, the estimated SCGF is calculated by interpolating all the values obtained.

Generally speaking, a direct implementation of the numerical scheme presented above, although extremely efficient from a computational point of view, could still lead to noisy results in the calculation of the dominant eigenvalue and right eigenvector in absence of good initial guesses for $r_s$ and $\zeta_s$. In the simulations used for Fig.\ \deletetext{(2)}\newtext{(1)}MT and Fig.\ \deletetext{(3)}\newtext{(2)}MT we have initialized the vector $r_s$ with random components sampled from the $[0,1]$ uniform distribution and then normalized the vector according to the $L^1$ norm. We will detail this point in Section \ref{sec:opt}. We also address the reader to Chapter 4 of \cite{Coghi2021} for a more general discussion.

\subsubsection{Pseudocode}

For full transparency of the work done, we propose in the following a pseudocode for the ARW, obtained by implementing the algorithm presented in the previous Section \ref{sec:ARW} for the particular case of a transition matrix $\Pi$ associated with an URW and tilting parameter $s=1$. We address the reader to the GitHub folder\footnote{Codes available at \url{https://github.com/gabriele-di-bona/ARW}\newtext{.}} where we have collected all the codes used to draft the manuscript.

\begin{algorithm}
	\SetKwInOut{Input}{Input}
    \SetKwInOut{Output}{Output}

    \Input{$\Pi$, $N$ (actual size of the graph), $k$ (degree list), $\deletetext{\alpha}\newtexttwo{\beta}$ (learning rate exponent), $\text{Max}$ (maximum number of iterations)}
    \Output{right dominant eigenvector: $r_1$, dominant eigenvalue: $\zeta_1$}
    
    \tcp{Initialisation}
    
	$r_1 = $ $L^1$-norm rnd list \;
	$\rho = 0$ list initialisation of the (non-normalized) empirical occupation measure \;
	$\text{start} = $ rnd($N$) \;
	$\rho(\text{start}) \mathrel{{+}{=}} 1$ \;
	$i_0$ = rnd($N$) \;

    \For{$n \gets 0$ \textbf{to} \text{Max}}
    {

   		\tcp{setting the learning rate}
    	$\deletetext{a}\newtexttwo{\lambda}(n) = \frac{1}{(1+n)^{\deletetext{\alpha}\newtexttwo{\beta}}}$ \;
		\If{$n != 0$}
      	{	
        	start $=$ next \;
        	\tcp{Empirical occupation measure}
	        $\rho(\text{start}) \mathrel{{+}{=}} 1$ \;
        	\For{$j \gets 1$ \textbf{to} $N$}
        	{
      			\tcp{step 1: driven process update in Eq.\ (\ref{eq:DrivenEstimate}).}
      			$\pi^{(n)}_{\text{start}j} = \frac{\pi^{(n-1)}_{\text{start}j} r_1(j)}{\sum_{j'=1}^N \pi^{(n-1)}_{\text{start}j'} r_1(j')}$ \;
            }
      	}

      	\tcp{step 2: pick a random index according to the probability vector $\pi^{(n)}_{\text{start}}$ (biased random walk).}
      	next $=$ rnd($N$,$\pi^{(n)}_{\text{start}}$) \;

		\tcp{step 3: asynchronous update of the right eigenvector in Eq.\ (\ref{eq:RightEigvApproxAdaptLearn}).}        
        $r_1(\text{start}) = r_1(\text{start}) + \deletetext{a}\newtexttwo{\lambda}(n) \left( \frac{k(\text{start}) \sum_{j \in \partial \text{start} } r_1(j)}{k_{\text{start}} r_1(i_0)} - r_1(\text{start}) \right)$ \;
        
        $i_0 = \max(r_1)$ \;
    }
    \Return $r_1$, $\zeta_1 \coloneqq r_1(i_0)$
    \caption{Adaptive random walk (ARW)}
    \label{algo}
\end{algorithm}

\newtext{
\section{Details on the entropy production rate}

As we have stated in the previous Section, in a large deviation context the SCGF is fundamental to obtain the entropy production rate $h(s)$. 
The SCGF can be calculated starting from the tilted matrix $\tilde{\Pi}_s = \left\lbrace \left( \tilde{\pi}_s \right)_{ij} \right\rbrace$, which in the case shown in the MT has components
\begin{equation}
\label{eq:TiltedMatrix}
\left( \tilde{\pi}_s \right)_{ij} = \pi_{ij} e^{s \ln k_i} = \pi_{ij} k_i^s = a_{ij} k_i^{s-1} \ ,
\end{equation}
where we have used $f(i) = k_i$ as function of the state. 
Hence, the SCGF is given by the logarithm of the dominant eigenvalue $\zeta_s$ of $\tilde{\Pi}_s$, that is
\begin{equation}
    \label{eq:def_Psi}
    \Psi(s) = \log(\zeta_s) \ .
\end{equation}
Eventually, the entropy rate of the driven process can be obtained taking Eq.~(8)MT and plugging it into Eq.~(3)MT, and can be expressed in terms of the SCGF as in Eq.~(10)MT~\cite{Coghi2019}.

The entropy rate of the driven process $h(s)$ (given by Eq.~(10)MT) has a unique global maximum in $s=1$. This can be seen by taking the derivative of $h(s)$ with respect to $s$ and putting it equal to $0$. This reads
\begin{equation}
    \label{eq:MaxEntrRate}
    0 = (1-s)\Psi''(s) \ .
\end{equation}
We notice that if the SCGF $\Psi$ exists, then it is analytic and convex \cite{Touchette2009,Dembo2010}. Hence, we have that $\Psi''(s) \geq 0$. As a consequence, Eq.~\eqref{eq:MaxEntrRate} is satisfied either for $s=1$ or when $\Psi''(s)=0$. Since for finite graphs $\Psi''(s)=0$ can only be true for $s \rightarrow \pm \infty$ (otherwise the function would show some non-analyticities for finite values of $s$), we are left with only the critical point $s=1$, which can be directly checked to be a maximum of the function $h$.

In Fig.\ \ref{fig:SCGF}, we plot the entropy rate $h(s)$ for an Erd\"{o}s-R\'{e}nyi graph with 100 nodes and average degree $3$, computed using Eq.~\eqref{eq:def_Psi} and Eq.~(10)MT. 
\begin{figure}[h!]
\centering
\includegraphics[width=.5\columnwidth]{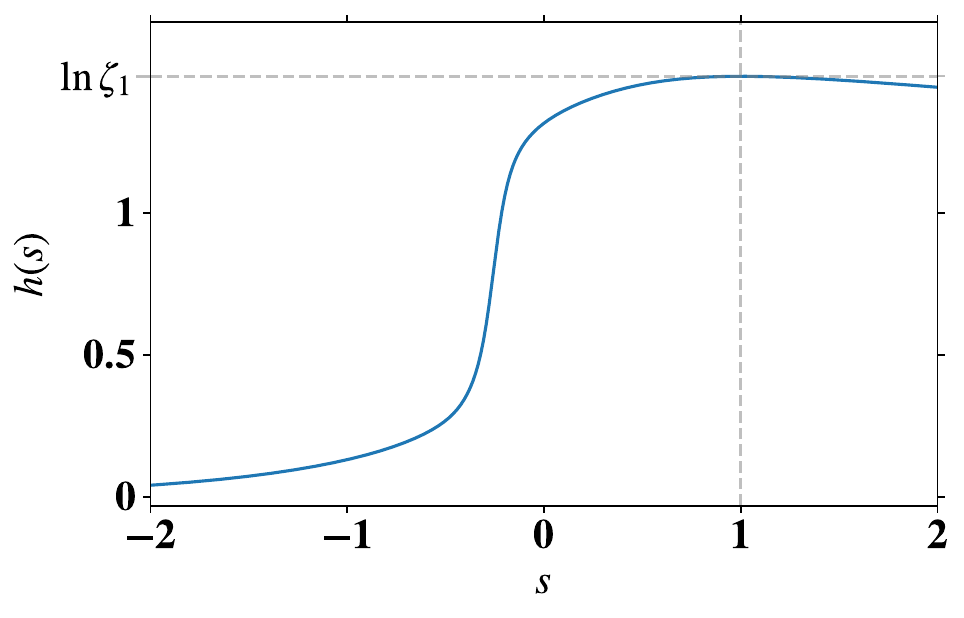}
\caption{\newtext{Entropy production rate $h(s)$ of the driven process in Eq.~(12)MT on an Erd\"{o}s-R\'{e}nyi graph with 100 nodes and average degree $3$. This shows a maximum for $s=1$.}}
\label{fig:SCGF}
\end{figure}

}

\newtexttwo{
\section{Setting off the learning rate for the adaptive random walk}
\label{sec:setlearning}

In this Section we give details on how to set off the learning rate $\lambda$ in Eq.\ \eqref{eq:LearningRate} appearing in the ARW algorithm (see Eq.\ \eqref{eq:RightEigvApproxAdaptLearn}). As we mentioned in Section\ \ref{sec:ARW}, an exact functional form of the learning rate $\lambda$ as a function of the network topology and the initial conditions does not exist in the literature and is, arguably, hard to find. Even harder---perhaps impossible---to find is a good (optimal), form of the learning rate without having any information on the actual exploration space. These are very much open questions that people in the field of stochastic approximation and reinforcement learning are pursuing. Nonetheless important, here, we do not seek to find an answer for them, and we limit ourselves to show a possible numerical way that one can implement to set off a good learning rate.

First of all, we make use of the general learning rate form of Eq.\ \eqref{eq:LearningRate} that was proposed and used in many other works related to stochastic approximation methods, see for instance \cite{Hsieh2002,Borkar2003,Ahamed2006}. Given this form, the problem boils down to finding a good exponent $\beta$.  Although this must satisfy the mathematical conditions stated in \cite{Borkar1998} to guarantee numerical converge, this requirement is not enough to select one single $\beta$ or a small set of them. Indeed, it is often the case that although the mathematical convergence is guaranteed, numerically one faces a full spectrum of different behaviour: too slow convergence, too fast convergence, numerical instability---all leading to bad learning. To avoid such a situation, a numerical tuning of $\beta$ appears necessary.

To set off a good value of $\beta$ that we have used in all our simulations with the most different topologies, we use Erd\"{o}s--R\'{e}nyi random graphs as toy models. As shown in the following figures, the fact that the same $\beta$, `optimally' set for Erd\"{o}s--R\'{e}nyi random graphs, appears to work as well also for scale-free random graphs is symptom of numerical stability. In fact, this numerical stability appears to be very pronounced as the same $\beta$ shows to work well even on Erd\"{o}s--R\'{e}nyi graphs of very different mean degree.

In Fig.\ \ref{fig:ARWBetaKAbs}, and \ref{fig:ARWBetaKAbsE}, we respectively analyse and plot how the entropy production rate calculated when $M \in \left\lbrace 100,\, 500,\, 1000 \right\rbrace$ or $M \in \left\lbrace 0.01E,\, 0.1E,\, 0.3E \right\rbrace$ links of a graph ($E$ is the size of the set of links in the graph, introduced to show that we do not need to know the exact number of links explored) have been explored by ARWs with different exponents $\beta \in \left\lbrace 0.01,\, 0.05,\, 0.1,\, 0.25,\, 0.5,\, 1.0 \right\rbrace$ varies with the mean degree of Erd\"{o}s--R\'{e}nyi and Barabasi--Albert random graphs. It appears evident that all values of $\beta \lesssim 0.25$ work very well, reaching a very high entropy production rate. We stress here that although we have not compared the entropy production rate of ARWs with the real optimal one given by the logarithm of the adjacency matrix of the $M$ links explored, by the compression of the lines with decreasing $\beta$ one may, correctly (see comparison with Fig.\ \ref{fig:ARWBetaKRel}, and \ref{fig:ARWBetaKRelE}), infer that those $\beta$s are the best ones, i.e., leading to a numerical convergence of the entropy production rate given by Eq.\ \eqref{eq:entrratevisitednodes} very close to the optimal one.

\begin{figure*}[t]
\centering
\includegraphics[width=\textwidth]{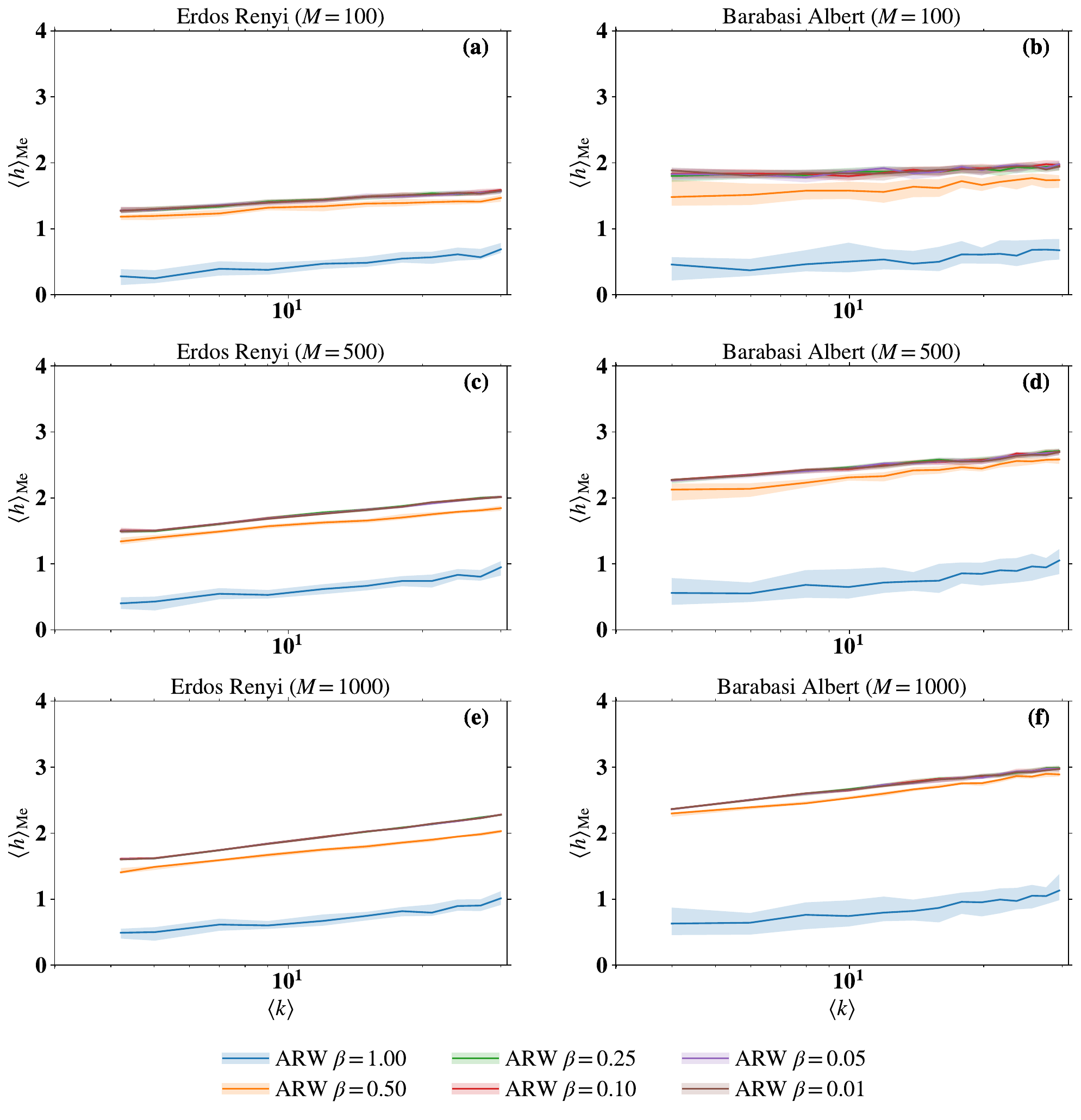}
\caption{\newtexttwo{(a,c,e)
The median entropy production rate $\left\langle h \right\rangle_{\text{Me}}$, calculated as in Eq.~\eqref{eq:entrratevisitednodes} when $M = 100$, $500$, $1000$ links have been explored, of $50$ trajectories of ARWs with different $\beta \in (0.01$, $0.05$, $0.1$, $0.25$, $0.5$, $1.0)$ as a function of the mean degree $\left\langle k \right\rangle = 4,5,7,9,12,15,18,21,24,27,30$ of an Erd\"{o}s--R\'{e}nyi random graph with $1000$ nodes. (b,d,f) Similarly to (a,c,e) but for a Barabasi--Albert random graph with $1000$ nodes and mean degree $\left\langle k \right\rangle = 4,6,8,10,12,14,16,18,20,22,24,26,28,30$. The shaded area around each solid line is delimited by the first and third quartiles.}}
\label{fig:ARWBetaKAbs}
\end{figure*}

\begin{figure*}[t]
\centering
\includegraphics[width=\textwidth]{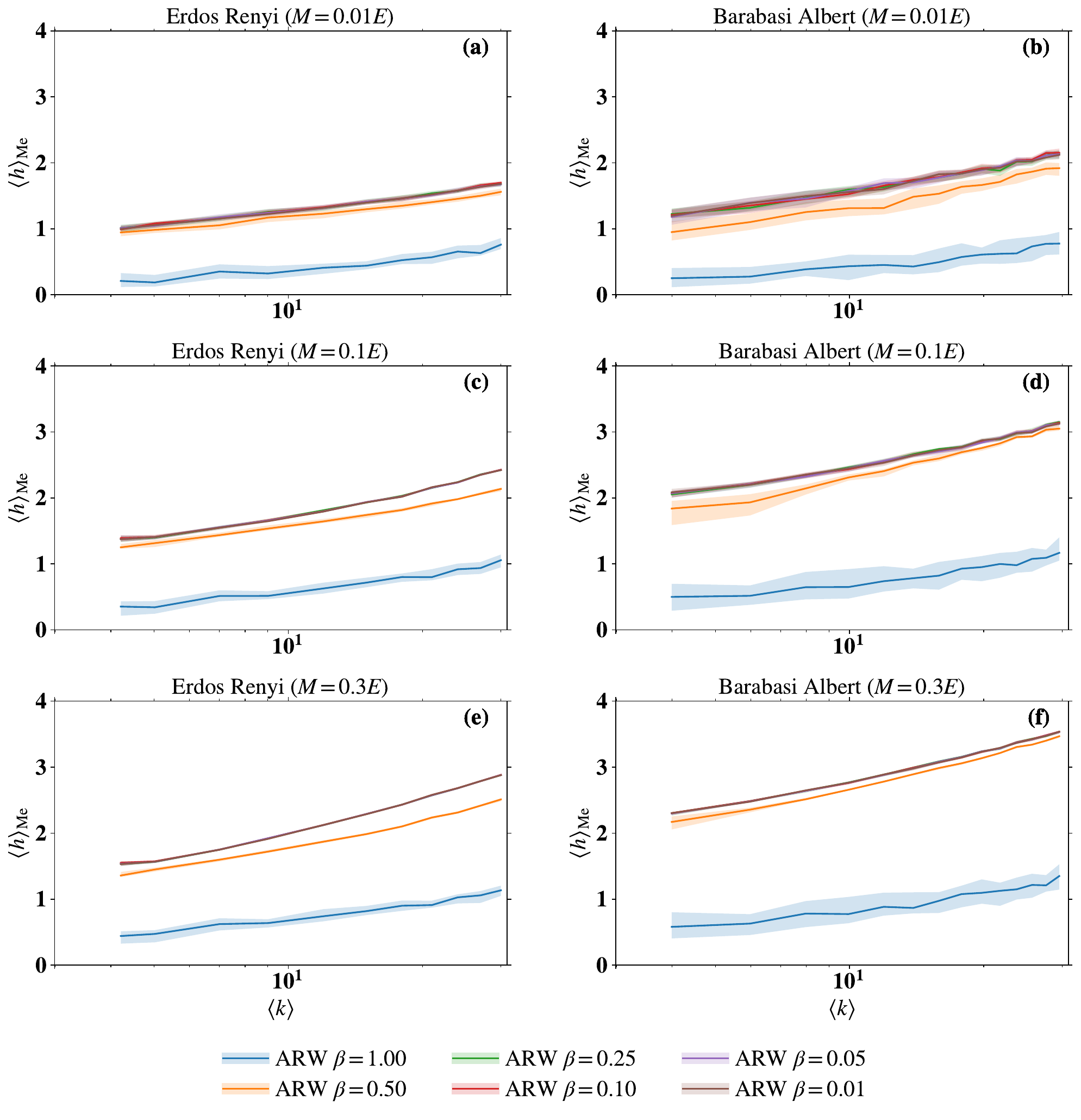}
\caption{\newtexttwo{(a,c,e)
The median entropy production rate $\left\langle h \right\rangle_{\text{Me}}$, calculated as in Eq.~\eqref{eq:entrratevisitednodes} when $M = 0.01E,0.1E,0.3E $ links have been explored, of $50$ trajectories of ARWs with different $\beta \in (0.01, 0.05, 0.1, 0.25, 0.5, 1.0)$ as a function of the mean degree $\left\langle k \right\rangle = 4,5,7,9,12,15,18,21,24,27,30$ of an Erd\"{o}s--R\'{e}nyi random graph with $1000$ nodes. (b,d,f) Similarly to (a,c,e) but for a Barabasi--Albert random graph with $1000$ nodes and mean degree $\left\langle k \right\rangle = 4,6,8,10,12,14,16,18,20,22,24,26,28,30$. The shaded area around each solid line is delimited by the first and third quartiles.}}
\label{fig:ARWBetaKAbsE}
\end{figure*}

In order to further demonstrate that the previous two figures already allow to infer the best $\beta$s, we plot in Fig.\ \ref{fig:ARWBetaKRel}, and \ref{fig:ARWBetaKRelE}, the relative error of the entropy production rate measured by \eqref{eq:entrratevisitednodes} with respect to the optimal one given by the logarithm of the adjacency matrix of the $M$ links explored. 

\begin{figure*}[t]
\centering
\includegraphics[width=\textwidth]{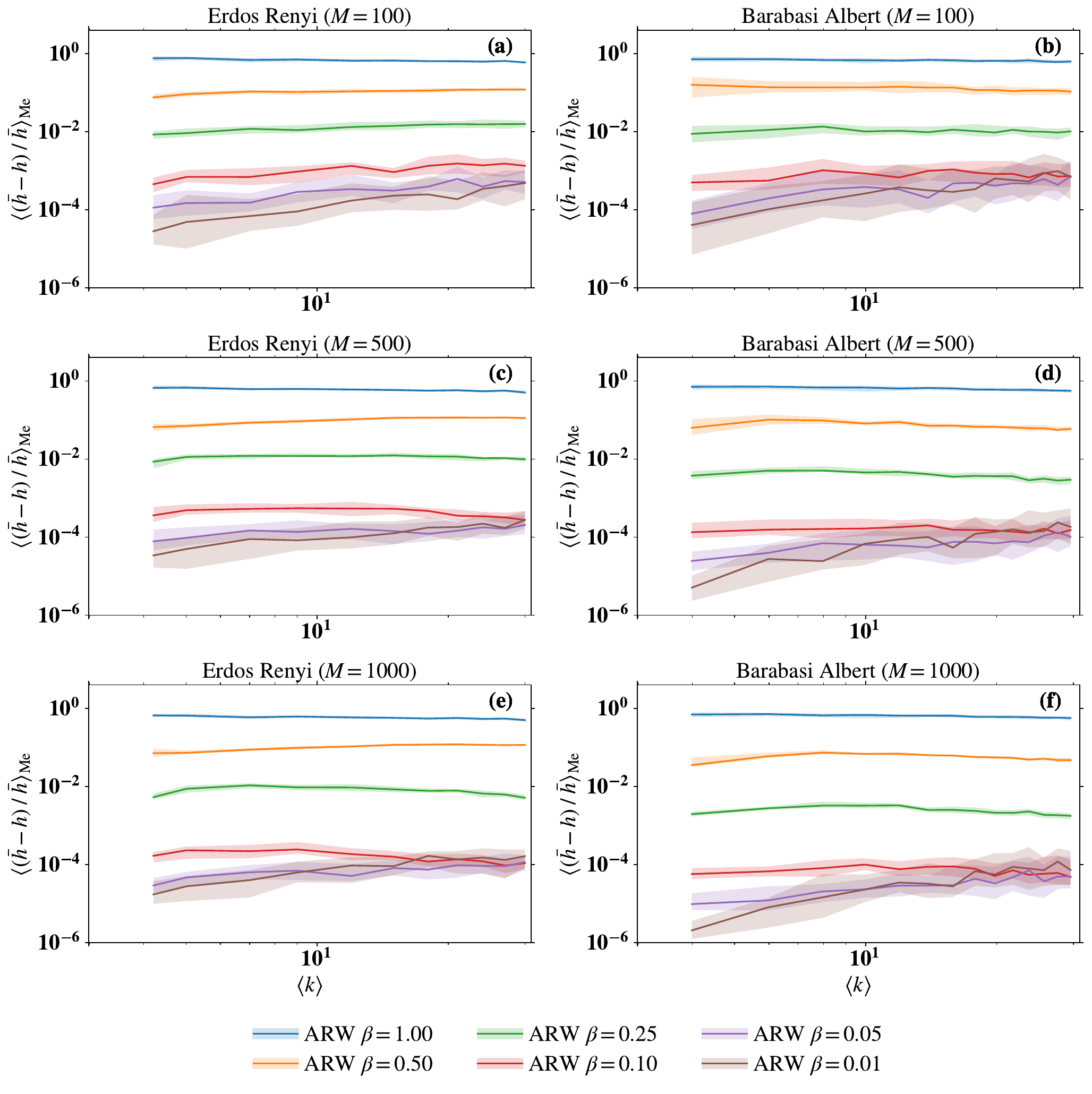}
\caption{\newtexttwo{(a,c,e)
The median relative error of the entropy production rate $\left\langle (\bar{h} - h)/\bar{h} \right\rangle_{\text{Me}}$, with $h$ calculated as in Eq.~\eqref{eq:entrratevisitednodes} when $M = 100, 500, 1000$ links have been explored, of $50$ trajectories of ARWs with different $\beta \in (0.01, 0.05, 0.1, 0.25, 0.5, 1.0)$ as a function of the mean degree $\left\langle k \right\rangle = 4,5,7,9,12,15,18,21,24,27,30$ of an Erd\"{o}s--R\'{e}nyi random graph with $1000$ nodes. (b,d,f) Similarly to (a,c,e) but for a Barabasi--Albert random graph with $1000$ nodes and mean degree $\left\langle k \right\rangle = 4,6,8,10,12,14,16,18,20,22,24,26,28,30$. The shaded area around each solid line is delimited by the first and third quartiles.}}
\label{fig:ARWBetaKRel}
\end{figure*}

\begin{figure*}[t]
\centering
\includegraphics[width=\textwidth]{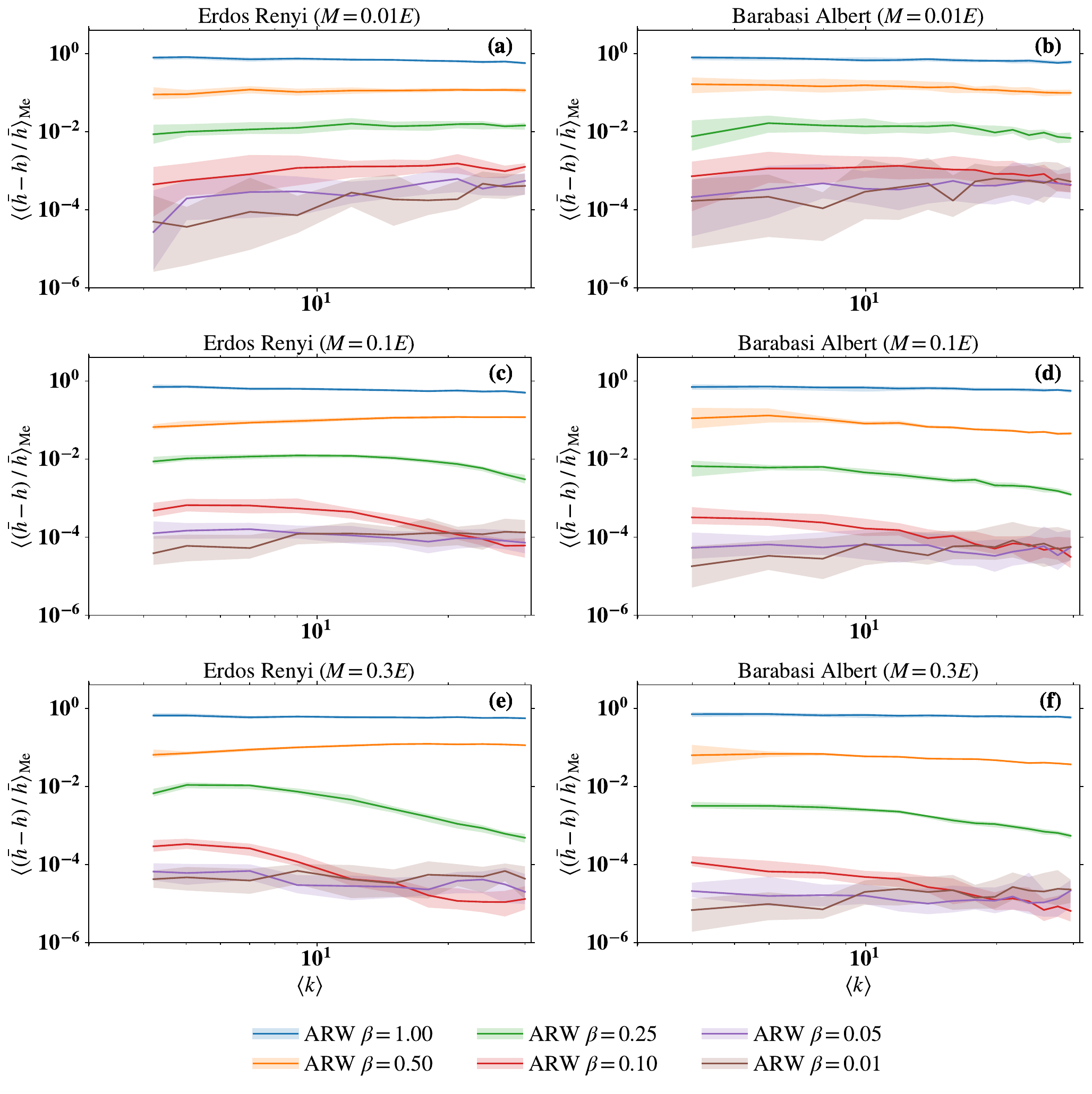}
\caption{\newtexttwo{(a,c,e)
The median relative error of the entropy production rate $\left\langle (\bar{h} - h)/\bar{h} \right\rangle_{\text{Me}}$, with $h$ calculated as in Eq.~\eqref{eq:entrratevisitednodes} when $M = 0.01E,0.1E,0.3E $ links have been explored, of $50$ trajectories of ARWs with different $\beta \in (0.01, 0.05, 0.1, 0.25, 0.5, 1.0)$ as a function of the mean degree $\left\langle k \right\rangle = 4,5,7,9,12,15,18,21,24,27,30$ of an Erd\"{o}s--R\'{e}nyi random graph with $1000$ nodes. (b,d,f) Similarly to (a,c,e) but for a Barabasi--Albert random graph with $1000$ nodes and mean degree $\left\langle k \right\rangle = 4,6,8,10,12,14,16,18,20,22,24,26,28,30$. The shaded area around each solid line is delimited by the first and third quartiles.}}
\label{fig:ARWBetaKRelE}
\end{figure*}

From Fig.\ \ref{fig:ARWBetaKRel}, and \ref{fig:ARWBetaKRelE} two further interesting remarks are in order. Firstly, by the fact that lines are, on average, mostly flat with varying $\left\langle k \right\rangle$ we conclude that the relative error of the entropy production rate does not really depend on the mean degree. Therefore, we can test the best $\beta$ on a single Erd\"{o}s--R\'{e}nyi random graph with a certain number of nodes and average degree and use that result for any other graph. Secondly, it indeed seems the case that any value of $\beta \lesssim 0.25$ can be used to get an optimal entropy production rate. However, ARW algorithms with these $\beta$s are not all numerically stable the same way, as we show in Tab.\ \ref{tab}. Indeed, it turns out that for values of $\beta$ that are too small, the algorithms do not finish the simulation. Eventually, we conclude that the value $\beta = 0.1$ is a good trade-off between numerical stability and convergence properties to the optimal entropy production rate.

To conclude, we show another extended numerical study analysing the behaviour of the entropy production rate, and its relative error with respect to the optimal value, as a function of $\beta$ out of single trajectories, starting at random positions, that cover only a very small portion of $M=100$ or $M=1000$ links of big networks of $10000$ nodes. We plot these results in Fig.\ \ref{fig:ARWrandomBeta}.

\begin{figure*}[t]
\centering
\includegraphics[width=\textwidth]{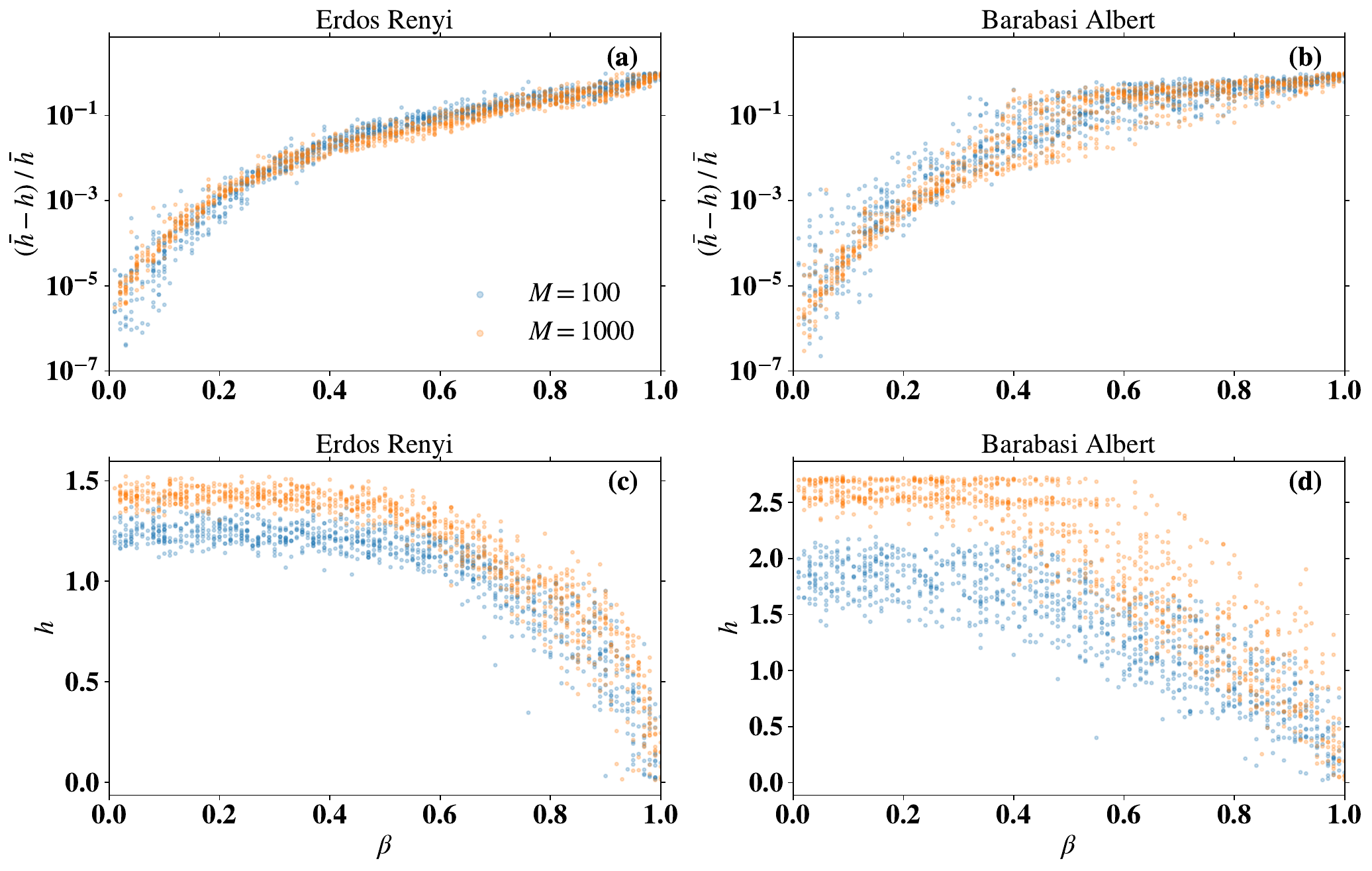}
\caption{\newtexttwo{(a,c) Respectively, the relative error of the entropy production rate $(\bar{h} - h)/\bar{h}$, and the entropy rate itself $h$ calculated as in Eq.~\eqref{eq:entrratevisitednodes} when $M = 100,1000$ links have been explored, of single trajectories of ARWs as a function of $\beta \in [0,1]$ for an Erd\"{o}s--R\'{e}nyi random graph with $10000$ nodes and mean degree $4$. (b,d) Similarly to (a,c) but for a Barabasi--Albert random graph with same size and mean degree. The shaded area around each solid line is delimited by the first and third quartiles.}}
\label{fig:ARWrandomBeta}
\end{figure*}

By investigating Figs.\ \ref{fig:ARWrandomBeta} (c) and (d) it is evident, once again, that the most interesting values one should consider in a simulation are characterised by a plateau for $\beta \lesssim 0.4$. Larger values lead to a smaller entropy production rate. Notice also that since the entropy is an extensive observable, the larger the number of links visited, the bigger the value of the entropy rate. Figs.\ \ref{fig:ARWrandomBeta} (a) and (b) are not affected by the size of the set of links visited and give further insights: the smaller the value of $\beta$, the smaller the relative error of the entropy production rate, but also the bigger the fluctuations that may be related to the numerical instability we previously discussed.
}

\clearpage

\section{Measuring the entropy production rate}
\label{sec:entrrate}

In Fig.\ (1)MT we compare the spreading performances of ARW, MERW, and URW by calculating the entropy production rates of each process while they explore the network. Clearly, there is some freedom in the choice of what to consider as \textit{explored} up to a certain moment in the evolution of each process. The most conservative choice, which is implemented in the calculations appearing in Fig.\ (1)MT, is to consider the graph visited composed only by the links (and connecting nodes) that have actually been crossed by the process in consideration. This is the case we face in the reality when, for instance, a person takes choices leading them to unknown situations that, in turn, are connected with other choices that, however, they do not know until the moment they face that particular situation. In such a scenario, the stochastic matrix entering in Eq.\ (17)MT at a certain time $n$ for each process will have to have non-zero entries only for the edges effectively crossed. This means that a redesign of the transition matrix---cancelling non-zero entries for unvisited links---for all the process while they run on the network needs to be done.

Another possibility is to consider the graph discovered composed by all the links that connect visited nodes. In the real world, this case happens every time one does not need to cross a link between two nodes to visit them if these have already been visited via other paths. Furthermore, a more theoretical reason pushing for this attempt comes from the implementation of the ARW as in the algorithm \ref{algo}: the right eigenvector is built on the nodes, not on the links, and is updated every time a node is visited, independently on the link used. Similarly, the transition matrix of the estimated driven process is updated for all the nodes connected to the one the random walk has just visited.  Similarly to the previous case, the transition matrices of ARW, MERW and URW need to be redesigned---cancelling non-zero entries for links that connect at least one unvisited node---every time a new node in the network is discovered. Therefore, in this scenario, the entropy production rate is calculated as
\begin{equation}
\label{eq:entrratevisitednodes}
h(M) = - \sum_{i,j \in V(n)} \rho_i(M) \pi_{ij}(M) \ln \pi_{ij}(M) \ ,
\end{equation}
where $V(n)$ is the set of visited nodes up to time $n$, and with a little abuse of notation $\pi_{ij}(M) = \pi_{ij}/ ( \sum_{i,j' \in V(n)} \pi_{ij'})$ if $(i,j) \in V(n)$, $0$ otherwise, and $\rho(M)$ is calculated as the dominant left eigenvector of $\Pi(M)$ as in the case treated in the MT.

\begin{figure*}[hbt]
\centering
\includegraphics[width=\textwidth]{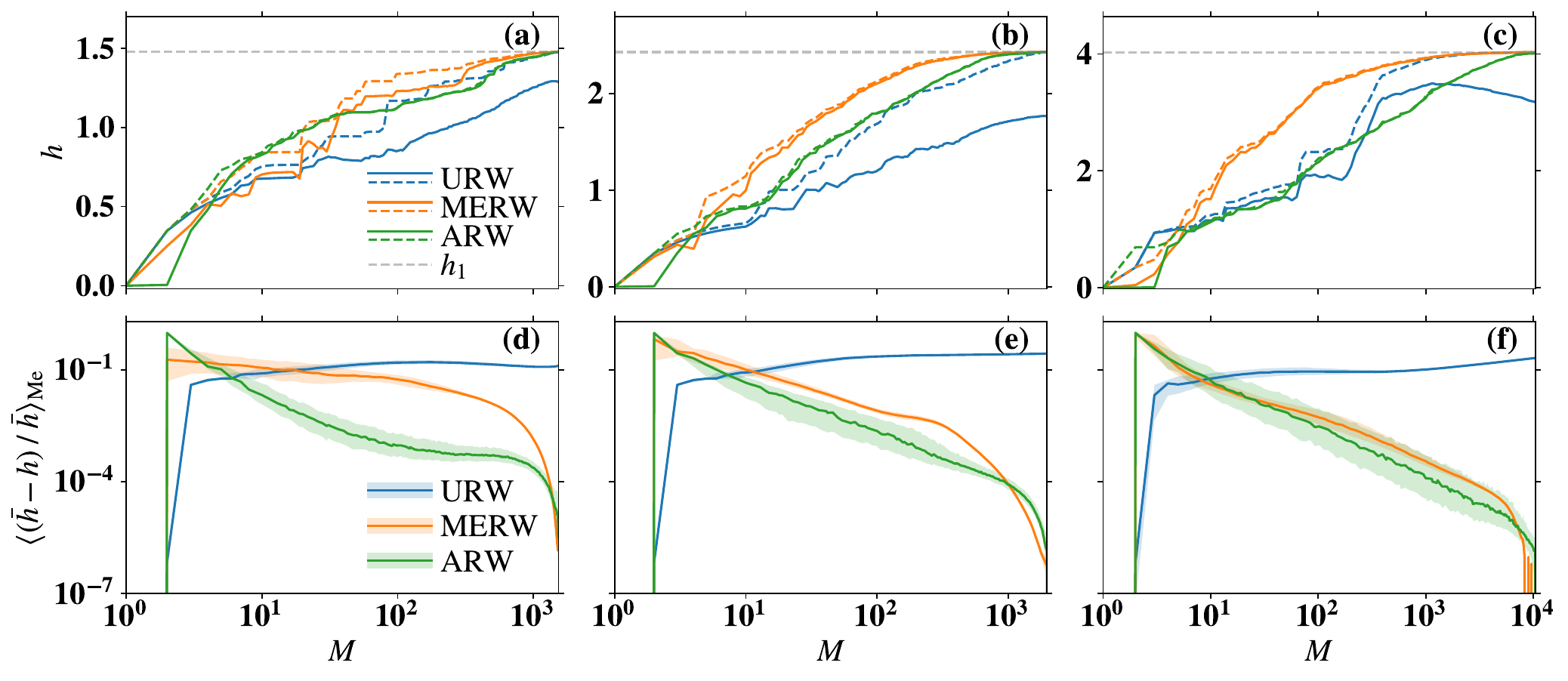}
\caption{(a-c)
The entropy production rate $h$, calculated as in Eq.~\eqref{eq:entrratevisitednodes}, of single trajectories of ARW, MERW and URW (solid lines) is compared to the corresponding optimal entropy production rate $\bar{h}$ on the discovered graph (dashed lines).
(d-f) Median (solid lines) and first and third quartiles (shaded area) of the normalized differences $(\bar{h}-h)/\bar{h}$ over an ensemble of $1000$ trajectories of ARW, MERW, and URW. 
Results are shown for random walks running on the giant connected component of an Erd\"{o}s--R\'{e}nyi random graph with $1000$ nodes and average degree $3$ in (a,d), a Barabasi--Albert network with $1000$ nodes and $m=\newtext{2}$\deletetext{4} in (b,e), and an air transportation network with $3618$ nodes and $14142$ links in (c,f).}
\label{fig:ARWMERWURWNodes}
\end{figure*}

In Fig.\ \ref{fig:ARWMERWURWNodes} we re-propose Fig.\ (1)MT for the entropy production rate calculated with Eq.~\eqref{eq:entrratevisitednodes}.
Noticeably, the ARW, obtained with a learning rate $\deletetext{a}\newtexttwo{\lambda}(n)=1/((n+1)^{0.1})$ in Eq.\ (14)MT or Eq.~\eqref{eq:RightEigvApproxAdaptLearn} for $s=1$, is only marginally better than MERW and URW on the Barabasi--Albert and on the air transportation network (the same analyzed in the MT), although still much better over the Erd\"{o}s--R\'{e}nyi random graph. This is evidence of the fact that the ARW is effectively optimized only over the links it actually crosses. For this reason, we argue that the entropy production rate needs to be calculated over the graph of visited links as done in Fig.\ (1)MT.

\section{Link-exploration time}
\label{sec:linktime}

\begin{figure*}[h]
\centering
\includegraphics[width=\textwidth]{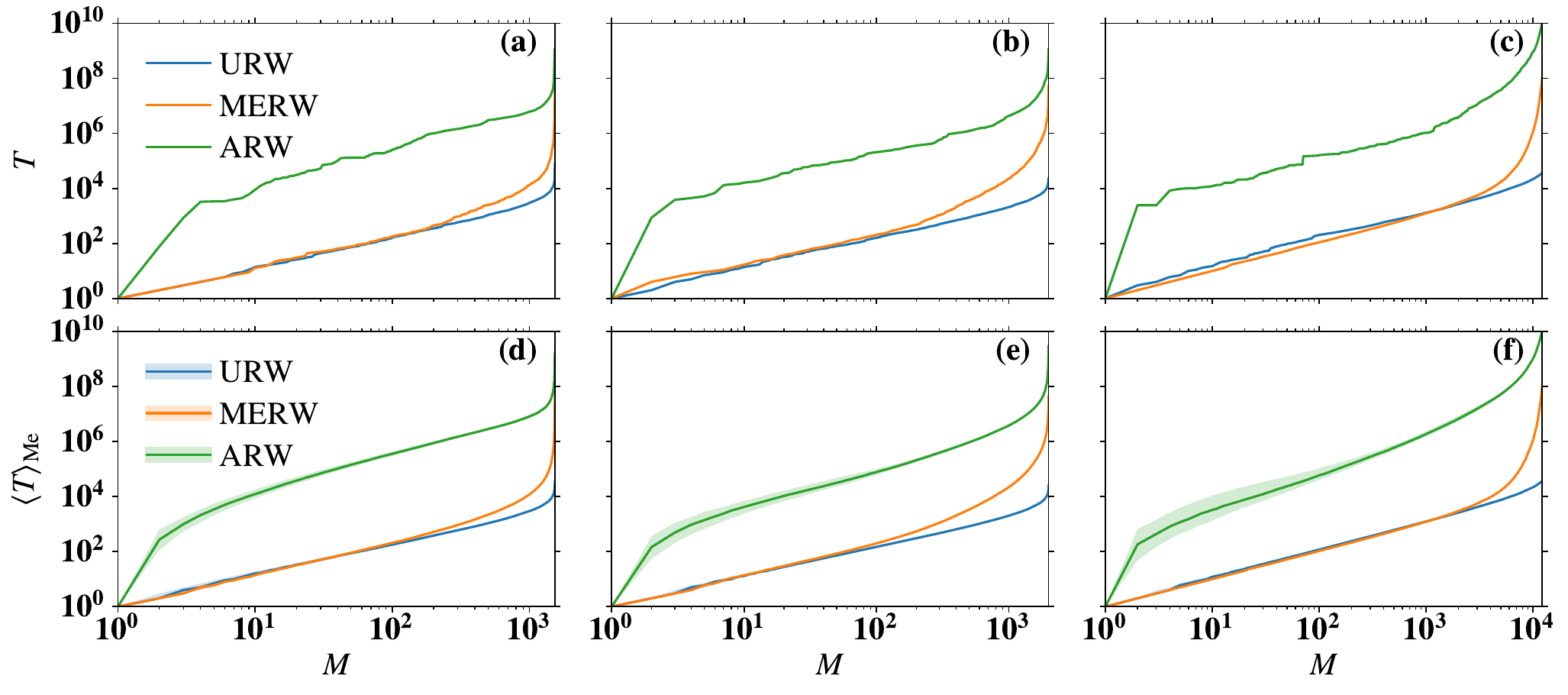}
\caption{(a-c) Link exploration time $T$ (solid line) for single trajectories of ARW, MERW, and URW. (d-f) Median (solid line), with first and third quartiles (shaded area), of the link exploration time for $1000$ trajectories. Results are shown for random walks running on the giant connected component of an Erd\"{o}s--R\'{e}nyi random graph of $1000$ nodes and average degree $3$ in (a) and (d), a Barabasi--Albert network of $1000$ nodes and $m=$\deletetext{4}\newtext{2} in (b) and (e), and an air transportation network with $3618$ nodes and $14142$ links in (c) and (f).}
\label{fig:ExplorationTime}
\end{figure*}

As mentioned in the MT, the ARW is characterized by a long warm-up time that prolongs the time to explore other links in the network. During the warm up, the ARW is localized in very few links (1 to 3) and by continuously crossing them it finely tunes the eigenvector centrality and the transition matrix estimate of the driven process in the visited portion of the graph, so as to set off an optimal exploration of the network. Indeed, once the warm up is over, the coverage time scaling exponent of the ARW is very similar to that of the MERW (see Fig.\ \deletetext{3 }\newtext{2}MT). In Fig.\ \ref{fig:ExplorationTime} we compare the link exploration time of MERW, URW, and ARW over the same networks considered in Fig.\ \deletetext{2 }\newtext{1}MT and in Fig.\ \ref{fig:ARWMERWURWNodes}. For each simulation, we save every time step in which the random walk explores a new link for the first time, and we plot them for a random instance of each process in Fig.\ \ref{fig:ExplorationTime}(a-c), while in Fig.\ \ref{fig:ExplorationTime}(d-f) we plot the medians, first and third quartiles, taken over $1000$ trajectories. Differently from Fig.\ \deletetext{3 }\newtext{2}MT, here we can analyze how the discovery rate of the process decreases or increases while the network is discovered highlighting the temporal extension of the warm-up time as well as the number of links involved.
As mentioned, this warm-up time is responsible for the longer---of about $10^2$ to $10^4$ steps---link coverage time of the ARW with respect to MERW and URW.

\section{Ideas on how to optimize the Adaptive Random Walk}
\label{sec:opt}

Such a long warm-up time seems to be necessary for the ARW to set off optimal initial conditions (in terms of right eigenvector and driven process transition matrix) for the exploration of the entire network. However, it also seems that the warm-up time can be shortened at the expense of the optimal spreading of the process. Indeed, the initial condition for the right eigenvector and the entropy production rate play as two conjugated quantities: the choice of the initial right eigenvector $r^{(0)}_1$ influences both the spreading properties and the overall coverage time of the process. \newtext{We would like to stress here that, conversely, the learning rate $\deletetext{a}\newtexttwo{\lambda}(n)$ appearing in the algorithm in Eq.~\eqref{eq:RightEigvApproxAdaptLearn} plays only a little role in determining the full coverage time of the network. Rather, it is pivotal in characterising the convergence of the algorithm in Eq.~\eqref{eq:RightEigvApproxAdaptLearn}, determining how high---maximal/optimal in the best case scenario---is the entropy production rate of the process obtained. Indeed, as we can see from Fig.\ \ref{fig:diff_a}, by setting off different processes with the most various values of $\deletetext{\alpha}\newtexttwo{\beta}$ in the learning rate in Eq.~\eqref{eq:LearningRate}, we show that all processes have, roughly, the same full coverage time of the network, although very different spreading (entropy production rate) propert\deletetext{y}\newtexttwo{ies}. We have run 1000 simulations for each process, with random initial node, on the giant connected component of an Erd\"{o}s--R\'{e}nyi random graph of $1000$ nodes and average degree $3$. 
\begin{figure*}[htb]
\centering
\includegraphics[width=\textwidth]{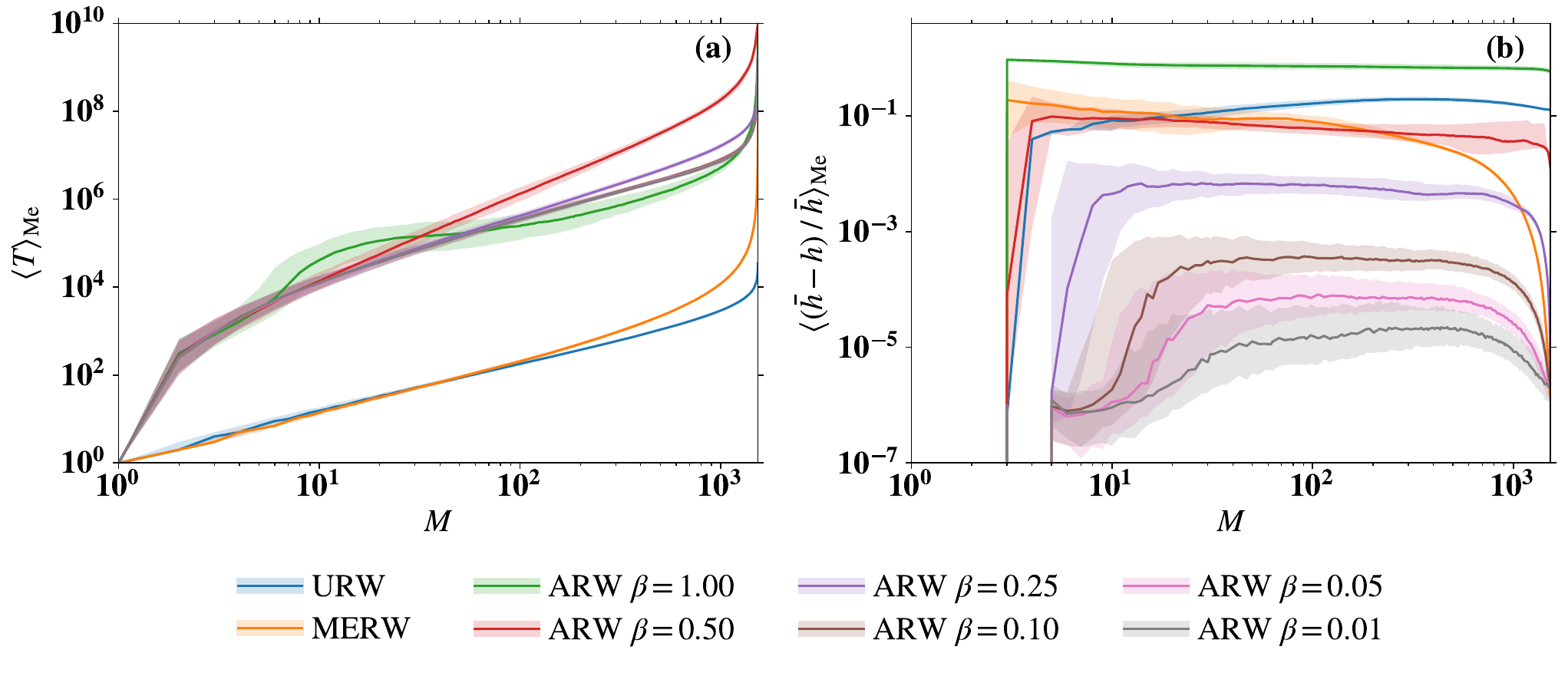}
\caption{Median (solid line), and first and third quartiles (shaded area), of the link exploration time (a) and of the entropy production rate relative difference (b) for $1000$ trajectories of URW, MERW, and ARW with different values of $\deletetext{\alpha}\newtexttwo{\beta}$ for $\deletetext{a}\newtexttwo{\lambda}(n)$ given in Eq.~\eqref{eq:LearningRate} (see legend below). Simulations run on the giant connected component of an Erd\"{o}s--R\'{e}nyi random graph of $1000$ nodes and average degree $3$.}
\label{fig:diff_a}
\end{figure*}
On the one hand, notice that in Fig.\ \ref{fig:diff_a}(a) the link exploration times of the ARW processes are almost all identical, higher than those of MERW and URW, while, in Fig.\ \ref{fig:diff_a}(b), the difference between the entropy production rate of the ARW process and the optimal one is much lower for lower values of $\deletetext{\alpha}\newtexttwo{\beta}$. Furthermore, notice that for $\deletetext{\alpha}\newtexttwo{\beta} \leq 0.25$ the ARW is much better than the MERW, while for higher values of $\deletetext{\alpha}\newtexttwo{\beta}$ the optimal convergence is not reached. In particular, in our simulations in the MT we keep $\deletetext{\alpha}\newtexttwo{\beta}=0.1$ as a good trade-off between numerical stability\newtexttwo{---see Tab.\ \ref{tab} where we compare for various choices of $\beta$ the number of times a simulation goes in overflow (\# Errors)---}and convergence to the maximal entropy production rate.} \newtext{As previously mentioned below Eq.~\eqref{eq:LearningRate}, we believe that this result can be further improved by analyzing more numerical simulations and various network topologies. All this, however, is beyond the scope of our paper.}

\begin{center}
\begin{table}
\newtexttwo{

\begin{tabular}[t]{l|lrr}

\toprule

Graph & Algorithm &  \# Simulations &  \# Errors \\

\hline

\midrule

Erd\"{o}s--R\'{e}nyi & URW & 1000 & 0 \\
Erd\"{o}s--R\'{e}nyi & MERW & 1000 & 0 \\
Erd\"{o}s--R\'{e}nyi & ARW $\beta=1.00$ & 1000 & 0 \\
Erd\"{o}s--R\'{e}nyi & ARW $\beta=0.50$ & 1000 & 0 \\
Erd\"{o}s--R\'{e}nyi & ARW $\beta=0.25$ & 1000 & 1 \\
Erd\"{o}s--R\'{e}nyi & ARW $\beta=0.10$ & 1000 & 0 \\
Erd\"{o}s--R\'{e}nyi & ARW $\beta=0.05$ & 1000 & 8 \\
Erd\"{o}s--R\'{e}nyi & ARW $\beta=0.01$ & 1000 & 94 \\
\bottomrule
\hline
\end{tabular}
\caption{\newtexttwo{For the same Erd\"{o}s--R\'{e}nyi random graph of $1000$ nodes and average degree $3$ used in Fig.\ \ref{fig:diff_a} we compare the number of times the simulation has crashed (\# Errors) on a total of $1000$ simulations for each process considered, possibly because of an overflow. Evidently, the numerical stability of the algorithm behind ARW decreases the smaller the $\beta$.}}
\label{tab}}
\end{table}
\end{center}

\begin{figure*}[htb]
\centering
\includegraphics[width=\textwidth]{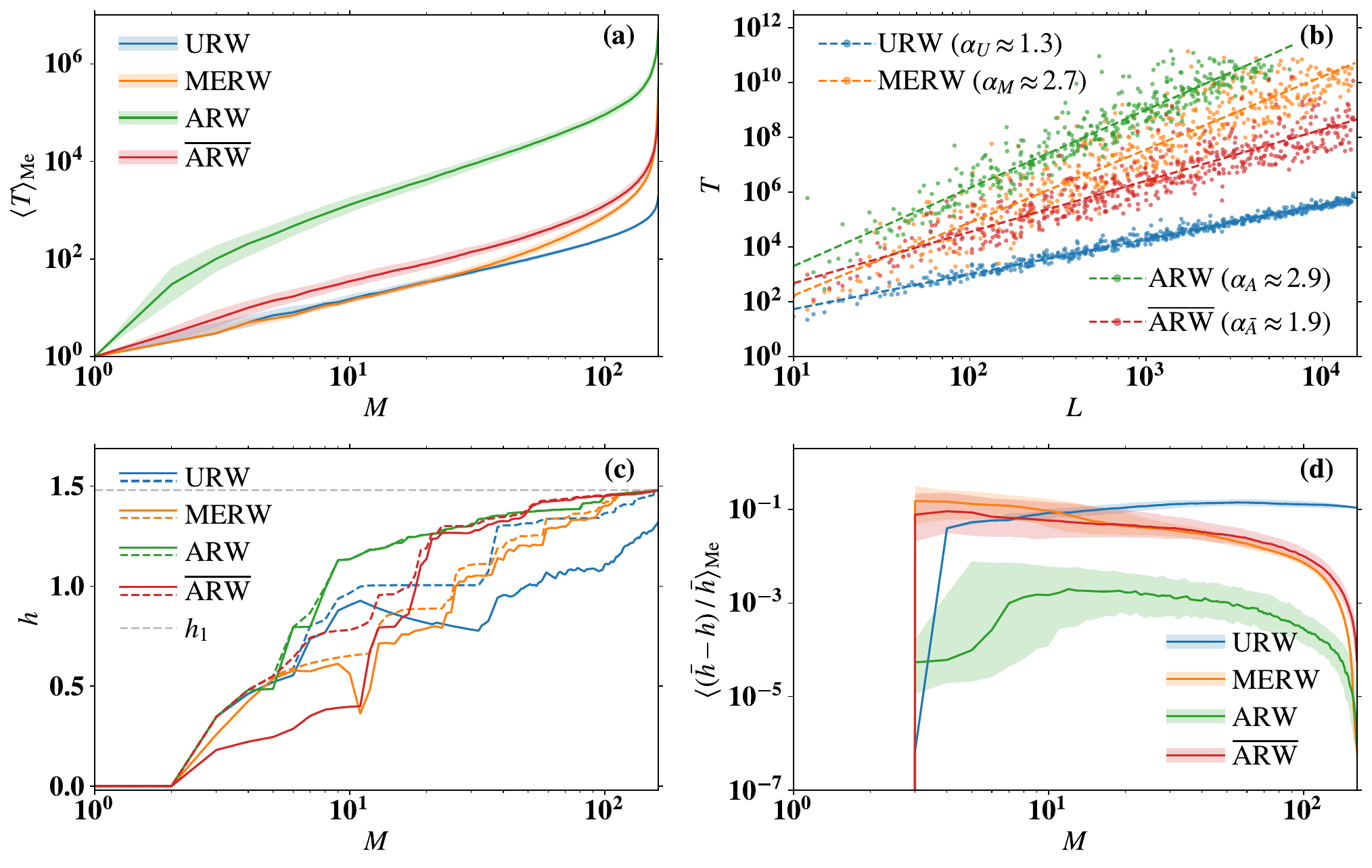}
\caption{(a) Median (solid line), and first and third quartiles (shaded area), of the link exploration time for $1000$ trajectories of URW, MERW, ARW, and $\overline{\text{ARW}}$ on the giant connected component of an Erd\"{o}s--R\'{e}nyi random graph of $100$ nodes and average degree $3$. (b) Coverage time $T$ as a function of $L$, the number of links in the giant connected component of different Erd\"{o}s-R\'{e}nyi graphs with average degree $3$ for URW, MERW, ARW, and $\overline{\text{ARW}}$. (c) Comparison between entropy production rate $h$ (solid line) and optimal entropy production rate $\bar{h}$ (dashed line) on the discovered graph, among single trajectories of URW, MERW, ARW, and $\overline{\text{ARW}}$. (d) Median (solid line), and first and third quartiles (shaded area), of the entropy production rate relative difference for $1000$ trajectories of URW, MERW, ARW, and $\overline{\text{ARW}}$.}
\label{fig:nonorm}
\end{figure*}

In the MT we considered $r^{(0)}_1$ to be an $L^1$-normalized random vector. In Fig.\ \ref{fig:nonorm} instead, we study the new $\overline{\text{ARW}}$, which is essentially equivalent to the ARW so far considered, apart from $r^{(0)}_1$ that, in this case, is a non-normalized random vector with components sampled from the uniform distribution over $[0,1]$.

Interestingly, $\overline{\text{ARW}}$ is much quicker than ARW in exploring the network---it has a smaller scaling exponent $\alpha_{\overline{\text{A}}}$---see Fig.\ \ref{fig:nonorm} (a) and (b) and it has about the same overall link coverage time of MERW. Furthermore, similarly to the MERW, it drastically decelerates in the last phase (Fig.\ \ref{fig:nonorm}(a)). Notwithstanding this, its spreading performance are, on average, no better than MERW (Fig.\ \ref{fig:nonorm}(c)(d)) and certainly not as good as ARW. 

This is only a preliminary study that shows an interesting interplay between initial conditions and overall spreading performances of the adaptive random walk. Further studies in this direction may help to find the best \newtext{conditions for the ARW} process \deletetext{that }\newtext{to} show\deletetext{s} short link exploring/coverage time and maximal entropy production rate while exploring the network.

\newpage

\bibliography{mybib}

\end{document}